\documentclass[a4paper,11pt]{article}
\usepackage{amssymb}
\usepackage{graphicx}
\usepackage{amsmath}
\usepackage{indentfirst}
\usepackage{mathrsfs}
\usepackage{graphicx}
\usepackage{cite}
\usepackage{hyperref}

\begin{document}


\title{BFV-Quantization of the Generalized Scalar Electrodynamics}

\author{R. Bufalo$^{1}$\thanks{%
rbufalo@ift.unesp.br}~, B.M. Pimentel$^{1}$\thanks{%
pimentel@ift.unesp.br}~ \\
\textit{{$^{1}${\small Instituto de F\'{\i}sica Te\'orica (IFT/UNESP), UNESP
- S\~ao Paulo State University}}} \\
\textit{\small Rua Dr. Bento Teobaldo Ferraz 271, Bloco II Barra Funda, CEP
01140-070 S\~ao Paulo, SP, Brazil}\\
}
\maketitle
\date{}

\begin{abstract}

This work comprises a study upon the quantization and the renormalizability of the generalized electrodynamics of spinless
charged particles (mesons), namely, the Generalized Scalar Electrodynamics ($GSQED_{4}$). The theory is quantized in the covariant framework
of the Batalin-Fradkin-Vilkovisky method. Thereafter, the complete Green's functions are obtained through functional methods
and a proper discussion on the theory's renormalizability is also given. Next, it is presented the computation and further
discussion on the radiative correction at $\alpha$-order; and, as it turns out, an unexpected $m_{P}$-dependent divergence on the mesonic sector
of the theory is found. Furthermore, in order to show the effectiveness of the renormalization
procedure on the present theory it is also shown a diagrammatic discussion on the photon self-energy at $\alpha^{2}$-order, where is observed
contributions from the meson self-energy function. Afterwards, we present the expressions of the counter-terms
and effective coupling of the theory. Obtaining from the later a energy range where the theory is defined $m^{2} \leq k^{2} < m_{P}^{2}$.
It is also shown in our final discussion that the new divergence is absorbed suitably by the mass counter-term $\delta_{Z_{0}}$, showing therefore that the
gauge WFT identities are satisfied still.

\end{abstract}

\maketitle


\section{INTRODUCTION}

\label{sec:1}

Higher-order derivative (HD) Lagrangians \cite{1} are a fairly interesting branch of
the ongoing effective theories \cite{2} in these thrilling times of the high-energy Physics we are living in. It is known that HD theories have,
as a field theory, better renormalization properties than the conventional ones. These properties had shown to be quite appealing in the
attempts to quantize gravity, where the Einstein action is supplied by terms containing higher powers of curvature leading to a renormalizable
\cite{3} and asymptotically free theory \cite{8}. Also, nowadays a new impetus in exploring appealing quantum theories of gravity has arisen,
for instance the $f(R)$-gravity \cite{4}, which is a strong candidate to explain the accelerating universe. Higher derivative theories were
proposed initially as an attempt to enhance and render a better ultraviolet behavior of physically relevant models \cite{5}. Once the effective
theory is rendered ultraviolet (UV) finite, we may consider it as an extension of the class of potentially interesting and consistent models
because its UV dynamics is now well defined. However, it was soon recognized that they possess a Hamiltonian which is not bounded from below
\cite{6} and that the process of adding such terms leads to the existence of negative norm states (or ghosts states) on the
quantum theory -- induces an indefinite metric in the space of states -- jeopardizing therefore the unitarity of the theory \cite{7}.
Despite the fact that many efforts have been made to overcome these ghost states, none of them was able to give a general method to
properly cope with this major problem \cite{8,9,10}. From the point of view of effective field theories a field theory violating unitarity
might still be sensitive at a low energy as long as the ghosts states are unstable so that they do not appear as asymptotic states.

As pointed out in several works \cite{11,12,13} along the years, it is long clear that Maxwell's theory is not the only one to describe
the Electromagnetic Field. One of the most successful generalization is the Generalized Electrodynamics \cite{11}. Actually, the Podolsky's
theory is the only one linear, Lorentz and $U(1)$ invariant generalization of Maxwell's theory \cite{13}. Another interesting feature inherent to the
Podolsky's theory is the existence of a generalized gauge condition also namely the generalized Lorenz condition: $\Omega \left[ A\right] =\left
( 1+m_{P}^{-2}\square \right) \partial ^{\mu }A_{\mu }$ \cite{14}; considered an important issue, once it is only through the
choice of the correct gauge condition that we can completely fix the gauge degrees of freedom of a given gauge theory \cite{14}. The authors with collaborators
have also been studying the Podolsky's theory at finite-temperature \cite{15,16} and some interesting results were obtained, such as implications
on the Podolsky's free parameter $m_{P}$.

In a series of previous works \cite{17,18}, we have presented the quantization and the renormalization of the Generalized Quantum Electrodynamics
($GQED_{4}$). The outcome in these works was encouraging as we observed that the theory's ultraviolet divergence was partially canceled at $\alpha$-order.
The electron self-energy and vertex part are both UV finite at $\alpha$-order only whether the generalized Lorenz condition is considered. Another
interesting point considered in the previous paper, it was the use of the experimental data of electron's anomalous magnetic
moment to bound possible values of the free parameter $m_{P}$; from such calculation it was found a consistent value as being: $m_{P}\geq3,7595\times10^{10} eV$.

In our opinion, the success of the theory of Generalized Quantum Electrodynamics warrants a similar investigation for other elementary particles. A first step
towards this direction is considered in the present paper, which is the study of the interaction of charged spinless mesons with the Podolsky electromagnetic field.
We shall denote this theory as the Generalized Scalar Quantum Electrodynamics ($GSQED_{4}$). Although a proper discussion on the Scalar Electrodynamics
features is a rare subject to find in the literature \cite{19}, there is an outstanding study of the Scalar Electrodynamics presented in rich details
and proofs in the Ref.\cite{20}. On theoretical grounds, $GSQED_{4}$ is an attractive theory, once its interaction Lagrangian contains
a richer structure than its fermionic counterpart. The use of spinless fields is fairly motivated in order to elucidate complicated
properties of a given theory, once it simplifies calculations in a theory due its spinless character and other nuances. Examples of that can be found
in investigations concerning $QCD$ properties \cite{21}, in which is believed that perturbative results (analytical results) can provide a natural
guide to possible nonperturbative structures.

Therefore, in this paper, we shall present a detailed study of the Generalized Scalar Quantum Electrodynamics
from a functional integral point of view. The main focus of the present discussion will be the analysis of the divergent behavior of $GSQED_{4}$.
The structure of the paper is as follows: In Sec.\ref{sec:2}, we present a study of the canonical structure of the theory by following Ostrogradski method to approach higher-derivative theories, and subsequently we construct the theory's transition-amplitude through the Batalin-Fradkin-Vilkovisky (BFV) procedure \cite{22}. Next, in the Sec.\ref{sec:3}, we introduce the
generating functional and, thereafter, we compute the fundamental Green's functions. Through this functional, we also derive the generalized
Ward-Fradkin-Takahashi (WFT) identities in Sec.\ref{sec:4}. In Sec.\ref{sec:5}, we apply to the $GSQED_{4}$ the on-shell renormalization
prescription and also discuss the appropriated renormalization conditions. In Sec.\ref{sec:6}, we evaluate the radiative corrections
of the theory at the $\alpha$-order approximation; discussing deeply the details of the divergent structure of the resulting expressions.
It is also presented a discussion on the divergent behavior of the photon self-energy function at $\alpha^{2}$-order.
Next, in Sec.\ref{sec:7}, we present the expressions for the counter-terms and for the effective coupling of the theory as well.
Our remarks and prospects are placed in the Sec.\ref{sec:8}. The Minkowski spacetime is concerned in the whole work, with the metric signature $(+,-,-,-)$.


\section{CONSTRAINT ANALYSIS AND TRANSITION-AMPLITUDE}

\label{sec:2}

It is well-known that effective theories may or may not be unitary. The unitarity is lost when a particle can lower
its energy by emiting other particles which have been eliminated in deriving the effective theory.
Actually there are sufficient reasons to dismiss any quantum theory, such as Einstein gravity,
that had the presence of HD quantum corrections and ghosts states. Motivated by the search of possible fundamental theories, one may naturally expect
the complete suppression of nonphysical, nonunitary processes. Nevertheless, in this paper we will assume that,
although we are not giving here a formal proof upon the details of physical space for Podolsky's theory, it could be performed an analysis,
for example through the generalized Kubo-Martin-Schwinger boundary conditions \cite{23}, or by the BRST symmetry and quartet mechanism \cite{24},
or even by the scheme proposed by Hawking and Hertog \cite{9}, leading therefore to a well-defined theory of the Generalized Electrodynamics.
In the following, we shall give a brief derivation on the constraint structure of $GSQED_{4}$ in order to construct the transition-amplitude
through the BFV method. Therefore, we have that the Lagrangian density describing the Generalized Scalar Electrodynamics is given by:
\begin{equation}
\mathcal{L}=\left( D_{\mu }\varphi \right) ^{\dag }D^{\mu }\varphi
-m^{2}\varphi ^{\dag }\varphi -\frac{1}{4}F_{\mu \nu }F^{\mu \nu }+\frac{1}{%
2m_{P}^{2}}\partial _{\mu }F^{\alpha \mu }\partial ^{\beta }F_{\alpha \beta
},  \label{eq 1.0}
\end{equation}%
where: $F_{\mu \nu }=\partial _{\nu }A_{\mu }-\partial _{\mu }A_{\nu }$ is
the usual electromagnetic field-strength tensor, and $D_{\mu }\varphi=\partial
_{\mu }\varphi-igA_{\mu }\varphi$ is the covariant derivative. The
interaction terms of the model Eq.\eqref{eq 1.0} can be written explicitly as:
\begin{eqnarray}
\mathcal{L}=\partial _{\mu }\varphi ^{\dag }\partial ^{\mu }\varphi
-m^{2}\varphi ^{\dag }\varphi +ig\varphi ^{\dag }\overleftrightarrow{%
\partial }^{\mu }\varphi A_{\mu }+g^{2}A_{\mu }\varphi ^{\dag }A^{\mu
}\varphi-\frac{1}{4}F_{\mu \nu }F^{\mu \nu }+\frac{1}{2m_{P}^{2}}\partial
_{\mu }F^{\alpha \mu }\partial ^{\beta }F_{\alpha \beta } .  \label{eq 1.1}
\end{eqnarray}%
Furthermore, the Lagrangian density \eqref{eq 1.0} is invariant, at the
classical level, under the local transformations:
\begin{equation}
\varphi \left( x\right) \rightarrow e^{ig\sigma \left( x\right) }\varphi
\left( x\right) ,\ A_{\mu }\rightarrow A_{\mu }+
\partial _{\mu }\sigma \left( x\right) .  \label{eq 1.2}
\end{equation}%
The Euler-Lagrange equations following from the Hamilton's principle \cite{12} are:
\begin{equation}
\left( \square +m^{2}\right) \varphi -ig\left( 2\partial _{\mu
}\varphi A^{\mu }+\varphi \partial _{\mu }A^{\mu }\right)
-g^{2}\eta^{\mu\nu}A_{\mu }A_{\nu }\varphi =0,  \label{eq 1.3}
\end{equation}%
and
\begin{equation}
\left( 1+m_{P}^{-2}\square \right) \partial _{\mu }F^{\alpha \mu }=j^{\alpha
} ,  \label{eq 1.4}
\end{equation}%
where: $j^{\alpha }=ig\varphi ^{\dag }\overleftrightarrow{\partial ^{\alpha }%
}\varphi +2g^{2}\varphi ^{\dag }A^{\alpha }\varphi $ is the scalar four-current.

In order to study the constraint structure of the present model, we must first
compute the canonical momenta of the field variables. Thus, the canonical momenta associated with the
scalar fields $\left( \varphi ,\varphi ^{\dag }\right) $ are:%
\begin{eqnarray}
\bar{p} &=&\frac{\partial \mathcal{L}}{\partial \left( \partial _{0}\varphi
^{\dag }\right) }=\partial _{0}\varphi -ig\varphi A_{0},  \notag \\
p &=&\frac{\partial \mathcal{L}}{\partial \left( \partial _{0}\varphi
\right) }=\partial _{0}\varphi ^{\dag }+ig\varphi ^{\dag }A_{0},
\label{eq 1.5}
\end{eqnarray}%
whereas the canonical momenta for gauge fields are obtained from Ostrogradski method to higher-order theories \cite{1}. This method consists in defining the dynamics of the system in a spanned phase space characterized by the independent variables: $\left( A_{\mu },\pi^{\nu}\right) $ and $\left( \Gamma_{\mu }\equiv \partial _{0}A_{\mu },\phi^\nu\right) $. Therefore, it follows that the generalized momenta associated with $\left( A_{\mu },\Gamma_{\mu }\right) $ \cite{14,25} are given by:%
\begin{eqnarray}
\phi ^{\alpha }&=& \frac{\partial \mathcal{L}}{\partial \left( \partial
_{0}\Gamma _{\alpha }\right) }=\frac{1}{m_{P}^{2}}\left( \eta ^{0\alpha
}\partial _{\lambda }F^{0\lambda }-\partial _{\mu }F^{\alpha \mu }\right) ,
\label{eq 1.6} \\
\pi ^{\mu } &=&\frac{\partial \mathcal{L}}{\partial \left( \Gamma_{\mu }\right) }-2\partial _{k}\left( \frac{\partial \mathcal{L}}{%
\partial \left( \partial _{k}\Gamma_{\mu }\right) }\right)
-\partial _{0}\left( \frac{\partial \mathcal{L}}{\partial \left( \partial
_{0}\Gamma_{\mu }\right) }\right)  \notag \\
&=&F^{\mu 0}-\frac{1}{m_{P}^{2}}\left( \eta ^{k\mu }\partial _{k}\partial
_{\lambda }F^{0\lambda }-\partial _{0}\partial _{\lambda }F^{\mu \lambda
}\right) .  \label{eq 1.7}
\end{eqnarray}%
From Eqs.\eqref{eq 1.6} and \eqref{eq 1.7}, and according to the linear
independence of the constraints \cite{25}, one obtains the following set of
first-class constraints:
\begin{eqnarray}
\Omega _{1}=\phi _{0}\approx 0,~~ \Omega _{2}=\pi _{0}-\partial ^{k}\phi
_{k}\approx 0, \notag \\
\Omega _{3}=\partial ^{k}\pi _{k}-ig\left( p\varphi -\bar{p}%
\varphi ^{\dag }\right) \approx 0,  \label{eq 1.8}
\end{eqnarray}
which comes strictly from the gauge sector, and none second-class constraint
is obtained. Here $\approx$ represents the fact that \eqref{eq 1.8} are
weak equations, according to Dirac's procedure \cite{25}.

As mentioned earlier, we will follow here the covariant framework of BFV method \cite{22} to
construct the transition-amplitude for the model:
\begin{eqnarray}
 Z &=&\int DA_{k}D\pi ^{k}D\Gamma _{l}D\phi ^{l}D\varphi ^{\dag }D\varphi DpD%
\bar{p}D\lambda DbD\bar{c}DcD\bar{P}DP   \label{eq 1.10}\\
&&\times \exp \Big[ i\int d^{4}x\Big[ \pi _{k}\dot{A}^{k}+\phi _{k}\dot{\Gamma} ^{k} +\left( \partial _{0}\varphi ^{\dag }\right)
\bar{p}+\dot{\varphi} p+\dot{c}\bar{P}  +\left( \partial _{0}\bar{c}\right) P+\dot{\lambda}b-\mathcal{H_{C}}\Big] +i\int
dw_{0}\left\{ \Psi ,Q_{BRST}\right\} \Big] , \notag
\end{eqnarray}%
with the quantities being defined by: $\left( c,\bar{P}\right) $ and $\left( \bar{c},P\right) $, the
ghost fields and its conjugated momenta, as $\left(\lambda,b\right) $ is a Lagrange
multiplier and its momentum, all satisfying the following Berezin brackets:
\begin{eqnarray}
\left\{ \bar{c}\left( z\right) ,P\left( w\right) \right\}_{B} =\delta \left(
z,w\right) ,\quad \left\{ \bar{P}\left( z\right) ,c\left( w\right)
\right\}_{B} =-\delta \left( z,w\right),\quad \left\{ \lambda \left( z\right) ,b\left( w\right) \right\}_{B}
=\delta \left( z,w\right) .
\end{eqnarray}
The canonical Hamiltonian $\mathcal{H_{C}}$ is given by:
\begin{eqnarray}
\mathcal{H}_{\mathcal{C}} &=&\pi _{0}\Gamma ^{0}+\pi _{j}\Gamma ^{j}+\frac{%
m_{P}^{2} }{2}\phi _{l}\phi ^{l}+\phi _{l}\partial ^{l}\Gamma _{0}+\phi
_{l}\partial _{k}F^{lk}+\bar{p}p +ig\left( p\varphi -\varphi ^{\dag }\bar{p}%
\right) A_{0}-\partial _{k}\varphi ^{\dag }\partial ^{k}\varphi
+m^{2}\varphi ^{\dag }\varphi  \notag \\
&& -ig\varphi ^{\dag }\overleftrightarrow{\partial ^{k}}\varphi
A_{k}-g^{2}A_{k}\varphi ^{\dag }A^{k}\varphi -\frac{1}{2}\left( \Gamma
_{j}-\partial _{j}A_{0}\right)^{2} +\frac{1}{4}F_{kj}F^{kj}-\frac{1}{%
2m_{P}^{2}}\left( \partial ^{j}\Gamma _{j}-\partial _{j}\partial
^{j}A_{0}\right) ^{2}.  \label{eq 1.11}
\end{eqnarray}%
Another quantity presents on \eqref{eq 1.10} is the BRST charge $Q$,
which has here the following form:
\begin{equation}
Q_{BRST}=\int d^{3}z\left[ c\left[ \partial _{k}\pi ^{k}-ig\left( p\varphi -%
\bar{p}\varphi ^{\dag }\right) \right] -iPb\right] .  \label{eq 1.12}
\end{equation}%
The quantity remaining to be defined here is the gauge-fixing function $%
\Psi$. Actually, one of the remarkable features of BFV method is that the
quantization procedure is independent of the choice of this function \cite{22}. However, we
will work here at the generalized radiation gauge condition:
\begin{eqnarray}
\Omega _{4}=A_{0}\approx 0,~~ \Omega _{5}=\Gamma _{0}\approx 0, \notag \\
\Omega _{6}=\left( 1+m_{P}^{-2}\square \right) \partial ^{k}A_{k}\approx
0, \label{eq 1.9}
\end{eqnarray}
which hence it allow us to write:
\begin{eqnarray}
\Psi =\int d^{3}w\Big[ \frac{i\xi }{2}b\bar{c}+i\bar{c}\left(
1+m_{P}^{-2}\square \right) \partial _{k}A^{k} -\lambda \left( 1+m_{P}^{-2}\square \right) ^{-1}\bar{P}\Big].  \label{eq 1.13}
\end{eqnarray}
Therefore, by computing $\left\{ \Psi ,Q_{BRST}\right\}$ and substituting
its resulting expression into \eqref{eq 1.10}, and then carrying out the momenta and
field variables integral, one finds the following expression for the
transition-amplitude:
\begin{eqnarray}
 Z &=&\int DA_{\mu }D\varphi ^{\dag }D\varphi D\bar{c}Dc  \exp \bigg\{i\int
d^{4}x\Big[\left( D_{\mu }\varphi \right) ^{\dag }D^{\mu }\varphi -m^{2}\varphi ^{\dag }\varphi  -\frac{1}{4}F_{\mu \nu }F^{\mu \nu }\notag \\
&&+\frac{1}{2m_{P}^{2}}\partial _{\mu
}F^{\mu \nu }\partial ^{\sigma }F_{\sigma \nu }-\frac{1}{2\xi }\left[ \left(1+m_{P}^{-2}\square \right) \partial _{\mu }A^{\mu }\right] ^{2} 
 +i\partial _{\mu }\bar{c}\left( 1+m_{P}^{-2}\square \right) \partial ^{\mu }c\Big]%
\bigg\}. \label{eq 1.15}
\end{eqnarray}%
Hence, from the BFV formalism we have obtained directly the desirable covariant expression
for the transition-amplitude. Furthermore, we see that the ghosts fields are decoupled from the
gauge fields, and so, their contribution can be absorbed into a normalization constant.


\section{SCHWINGER-DYSON-FRADKIN EQUATIONS}

\label{sec:3}

In the present section, we continue the formal development of the $GSQED_{4}$,
by deriving now coupled relations between the fundamental Green's functions,
which are known as the Schwinger-Dyson-Fradkin Equations (SDFE) \cite{26}. We will
consider the theory's fundamental Green's functions: the gauge field propagator $%
\mathcal{D}$, the meson field propagator $\mathcal{S}$ and the two vertex
functions, the three-point function $\Gamma_{\sigma}$ and the four-point
function $\Phi_{\sigma\rho}$. For the derivation of these quantities we will make use of the
functional methods, which provide a rather natural way to obtain such
functions. In this context, we need to define the generating functional:
\begin{equation}
\mathcal{Z}\left[ \zeta ,\bar{\zeta},J_{\mu }\right] =\int D\mu
\left(\varphi,\varphi^{\dag},A_{\mu }\right) \exp \left[ i\mathcal{A}\right]
,  \label{eq 2.0}
\end{equation}
with the action $\mathcal{A}$ given by:
\begin{eqnarray}
\mathcal{A} &=&\int d^{4}x\Big[\left(D_{\mu }\varphi \right) ^{\dag }D^{\mu }\varphi  -m^{2}\varphi ^{\dag }\varphi  -\frac{1}{4}F_{\mu \nu }F^{\mu \nu }  \notag \\
&&+\frac{1}{2m_{P}^{2}}\partial ^{\mu }F_{\mu \beta }\partial _{\alpha
}F^{\alpha \beta }-\frac{1}{2\xi }\left[ \left( 1+m_{P}^{-2}\square \right)
\partial ^{\mu }A_{\mu }\right] ^{2} +\bar{\zeta}\varphi +\varphi ^{\dag }\zeta +J_{\mu }A^{\mu }\Big],  \notag\\
&=&\mathcal{A}_{eff}+\int d^{4}x\Big[\bar{\zeta}\varphi +\varphi ^{\dag
}\zeta +J_{\mu }A^{\mu }\Big]. \label{eq 2.1}
\end{eqnarray}%
The sources $\zeta$, $\bar{\zeta}$ and $J_{\mu}$ are related with the
fields $\varphi^{\dag}$, $\varphi$ and $A_{\mu}$, respectively.


\subsection{SDFE for the photon propagator}

\label{sec:3.1}

We will derive here the complete expression of the photon
propagator. In order to obtain its SDFE equation, we need to solve:
\begin{equation}
\left[ \left. \frac{\delta \mathcal{A}_{eff}}{\delta A_{\mu }\left(
x\right) }\right\vert _{\frac{\delta }{i\delta \bar{\zeta}},\frac{\delta }{%
i\delta \zeta },\frac{\delta }{i\delta J_{\mu }}}+J^{\mu }\left( x\right) %
\right] \mathcal{Z}\left[ \zeta ,\bar{\zeta},J_{\mu }\right] =0,
\label{eq 2.2}
\end{equation}%
which, when used \eqref{eq 2.1} for the action $\mathcal{A}_{eff}$, can be rewritten as:
\begin{eqnarray}
-iJ^{\mu }\left( x\right) \mathcal{Z}=D^{\mu \nu }\frac{\delta \mathcal{Z}}{\delta
J^{\nu }\left( x\right) } +g\underset{z\rightarrow x}{\lim }\left[
\partial _{x}^{\mu }-\partial _{z}^{\mu }\right] \frac{\delta ^{2}\mathcal{Z}}{\delta \zeta
\left( z\right) \delta \bar{\zeta}\left( x\right) } -2g^{2}\frac{\delta^{3} \mathcal{Z}}{\delta
J_{\mu }\left( x\right)\delta \zeta \left( x\right) \delta \bar{\zeta}\left( x\right) }%
 ;  \label{eq 2.5}
\end{eqnarray}
where we have already defined the differential operator $D$ as the following:
\begin{equation}
D^{\mu \nu }=\left[ \square T^{\mu \nu }+\frac{1}{\xi }\left(
1+m_{P}^{-2}\square \right) \square L^{\mu \nu }\right] \left(
1+m_{P}^{-2}\square \right) ; \label{eq 2.3}
\end{equation}%
with the following set of differential projectors: $T^{\mu \nu }+L^{\mu \nu }=\eta ^{\mu
\nu };L^{\mu \nu }=\frac{\partial ^{\mu }\partial ^{\nu }}{\square }$.

It is interesting for our purposes to introduce new useful quantities related to the
generating functional $\mathcal{Z}$. Hereby, we first introduce the
generating functional for the connected Green's functions $W$, which is
defined by the relation: $W=-i\ln \mathcal{Z}$. It proves convenient to introduce also the generating
functional for the $1PI$ Green's functions, which is related to $W$ through
a Legendre transformation:
\begin{equation}
\Gamma \left[ \varphi ,\varphi ^{\dag },A_{\mu }\right] =W\left[ \bar{\zeta}%
,\zeta ,J_{\mu }\right] -\int d^{4}z\left[ \bar{\zeta}\varphi +\bar{\varphi}%
\zeta +J_{\mu }A^{\mu }\right] .  \label{eq 2.6}
\end{equation}%
From the above definitions, one can obtain identities relating the connected
and $1PI$ $2$-points functions. For instance, it follows that for the meson field:
\begin{equation}
\int d^{4}x\frac{\delta ^{2}\Gamma }{\delta \varphi \left( y\right) \delta
\bar{\varphi}\left( x\right) }\frac{\delta ^{2}W}{\delta \zeta \left(
x\right) \delta \bar{\zeta}\left( w\right) }=-\delta \left( w,y\right) .
\label{eq 2.8}
\end{equation}%
We also have the gauge field functionals satisfying a similar relation as the above.
Actually, these identities, such as \eqref{eq 2.8}, are an important key for
the formal development through functional methods and they will be used quite often throughout the work.
Anyhow, writing \eqref{eq 2.5} now in terms of $W$ and then differentiating the resulting
expression with respect to $J^{\nu }\left( y\right) $ it follows that:
\begin{eqnarray}
-\eta _{\nu }^{\mu }\delta \left( x,y\right) &=&D_{x}^{\mu \alpha }\frac{%
\delta ^{2}W}{\delta J^{\nu }\left( y\right) \delta J^{\alpha }\left(
x\right) }+g\underset{z\rightarrow x}{\lim }\left[ \partial _{x}^{\mu
}-\partial _{z}^{\mu }\right] \frac{\delta ^{3}W}{\delta J^{\nu }\left(
y\right) \delta \zeta \left( z\right) \delta \bar{\zeta}\left( x\right) }
\notag \\
&&-2g^{2}\underset{z\rightarrow x}{\lim }\bigg[ \frac{\delta ^{4}W}{\delta
J_{\mu }\left( x\right) \delta J^{\nu }\left( y\right) \delta \zeta \left(
z\right) \delta \bar{\zeta}\left( x\right) }+i\frac{\delta ^{2}W}{\delta
\zeta \left( z\right) \delta \bar{\zeta}\left( x\right) }\frac{\delta ^{2}W}{%
\delta J_{\mu }\left( x\right) \delta J^{\nu }\left( y\right) }\bigg] .
\label{eq 2.9}
\end{eqnarray}%
Now, defining a new functional quantity as:
\begin{eqnarray}
\Pi _{\mu \rho }\left( x,z\right) =-g\underset{h\rightarrow x}{\lim }\left[
\partial _{\mu }^{x}-\partial _{\mu }^{h}\right] \left( \Xi _{1}\right)
_{\rho }\left( x,h;z\right) +2g^{2}\underset{h\rightarrow x}{\lim }\left[
\left( \Xi _{2}\right) _{\mu \rho }\left( x,h;z\right) -i\eta _{\mu \rho
}\delta \left( x,z\right) \mathcal{S}\left(x,h\right)\right] ;  \label{eq 2.10}
\end{eqnarray}
with the quantities $\left( \Xi _{1}\right) _{\rho } $ and $\left( \Xi
_{2}\right) _{\rho } $ defined by \eqref{eq B.0} and \eqref{eq B.1},\footnote{In
order to avoid lengthy expressions in the body text, we present
some quantities, named as $\Xi_{j}$, in the Appendix \ref{sec:B}.}
respectively; also, by identifying $\mathcal{D}_{\sigma \mu }\left( w,x\right) $, Eq.\eqref{eq B.10}, as the
photon complete propagator, one obtains from \eqref{eq 2.9} the complete expression for the
propagator of the photon field in the Fourier space:
\begin{equation}
i\mathscr{D}_{\sigma \mu }=\frac{\eta _{\sigma \mu }-\frac{k_{\sigma }k_{\mu
}}{k^{2}}}{k^{2}\left[ \Pi \left( k\right) +\left( 1-m_{P}^{-2}k^{2}\right) %
\right] }+\frac{\xi }{k^{2}\left( 1-m_{P}^{-2}k^{2}\right) ^{2}}\frac{%
k_{\sigma }k_{\mu }}{k^{2}}.  \label{eq 2.12}
\end{equation}%
The functional $\Pi _{\mu \rho }$ is known as the polarization tensor, or
photon self-energy function, and it is related to the scalar polarization $%
\Pi $ through the structure:
\begin{equation}
\Pi ^{\mu \nu }\left( k\right) =\left( -\eta ^{\mu \nu }k^{2}+k^{\mu }k^{\nu
}\right) \Pi \left( k\right) .
\end{equation}%
Differently from the Generalized Quantum Electrodynamics \cite{17}, the
scalar version possesses new interaction terms and, consequently, a richer and rather
interesting structure on its radiative functions \eqref{eq 2.10}. We will present the evaluation
of the radiative correction functions at the lowest-order, and respective discussion on them, in the Sec.\ref{sec:6}.

Furthermore, the photon propagator at the lowest-order in perturbation
theory, can be conveniently written as:
\begin{eqnarray}
iD_{\sigma \mu }\left[ \eta _{\sigma \mu }-\left( 1-\xi \right) \frac{%
k_{\sigma }k_{\mu }}{k^{2}}\right] \frac{1}{k^{2}} -\left[ \eta _{\sigma \mu
}-\xi \frac{k_{\sigma }k_{\mu }}{k^{2}-m_{P}^{2}}\right] \frac{1}{%
k^{2}-m_{P}^{2}} +\left( 1-2\xi \right) \frac{k_{\sigma }k_{\mu }}{%
k^{2}\left( k^{2}-m_{P}^{2}\right) }.  \label{eq 2.15}
\end{eqnarray}
The above expression shows explicitly the contribution from the ghost states; and as said earlier, we will assume here that it
could be performed an analysis on that leading, hence, to a well-defined theory to the Generalized Electrodynamics. Also we are motivated
to retain some attention to the present theory, once its fermionic counterpart showed to be a rich theory, possessing a UV finite behavior
(at the light of effective theories) and interesting renormalized behavior.

Actually, the major feature of the structure of the photon free
propagator \eqref{eq 2.15} is observed when it is applied to radiative
correction calculation. As it was shown in a previous paper \cite{17}, the
contribution from the $m_{P}$-dependent terms act enhancing the UV behavior and
eliminating the UV divergences of some sectors of $GQED_{4}$.
However, as it was pointed out, this remarkable result is only attainable in the presence of a suitable
gauge condition, the generalized Lorenz condition \cite{7}. This point will be further discussed in the Sec.\ref{sec:6}.


\subsection{SDFE for the meson field propagator}

\label{sec:3.2}

In this subsection we will continue the derivation of the theory's SDFE, obtaining now an integral
expression for the complete meson propagator $\mathcal{S}$. We will follow
the guidelines presented previously on the derivation of the photon propagator.
Thereafter, we must first solve the following equation:
\begin{equation}
\left[ \left. \frac{\delta \mathcal{A}_{eff}}{\delta \varphi \left(
x\right) }\right\vert _{\frac{\delta }{i\delta \bar{\zeta}},\frac{\delta }{%
i\delta \zeta },\frac{\delta }{i\delta J_{\mu }}}+\bar{\zeta}\left( x\right) %
\right] \mathcal{Z}\left[ \zeta ,\bar{\zeta},J_{\mu }\right] =0.
\label{eq 2.16}
\end{equation}%
After a careful evaluation of the functional derivative of $\mathcal{A}_{eff}$ one can find the expression:%
\begin{eqnarray}
\left( \square +m^{2}\right) \frac{\delta \mathcal{Z} }{\delta \zeta \left( x\right) }%
+g\underset{z\rightarrow x}{\lim }\left[ 2\partial _{x}^{\mu
}+\partial _{z}^{\mu }\right] \frac{\delta ^{2}\mathcal{Z}}{\delta J^{\mu }\left( z\right) \delta \zeta \left( x\right)}+g^{2}\eta_{\mu\sigma}\frac{\delta ^{3}\mathcal{Z} }{\delta \zeta \left( x\right) \delta J_{\sigma }\left( x\right) \delta J_{\mu }\left( x\right) }=i\bar{\zeta}%
\left( x\right) \mathcal{Z}.  \label{eq 2.17}
\end{eqnarray}
Now, writing the above equation \eqref{eq 2.17}, in terms of the
generating functional $W$ and then differentiating the resulting expression with respect to the
source $\bar{\zeta}\left( y\right) $, one obtains:
\begin{eqnarray}
\delta \left( x,y\right) &=&\left( \square +m^{2}\right) \frac{\delta ^{2}W}{%
\delta \zeta \left( y\right) \delta \bar{\zeta}\left( x\right) }+g\underset{z\rightarrow x}{\lim }\left[ 2\partial _{y}^{\mu }+\partial _{z}^{\mu }%
\right] \frac{\delta ^{3}W}{\delta J^{\mu }\left( z\right) \delta \zeta
\left( y\right) \delta \bar{\zeta}\left( x\right) }  \notag \\
&&+g^{2}\eta _{\mu \sigma }\underset{z\rightarrow x}{\lim }\bigg[ \frac{%
\delta ^{4}W}{\delta J_{\mu }\left( x\right) \delta J_{\sigma }\left(
z\right) \delta \zeta \left( y\right) \delta \bar{\zeta}\left( x\right) }+i\frac{\delta ^{2}W}{\delta \zeta \left( y\right) \delta \bar{\zeta}\left(
x\right) }\frac{\delta ^{2}W}{\delta J_{\mu }\left( x\right) \delta
J_{\sigma }\left( z\right) }\bigg] .  \label{eq 2.18}
\end{eqnarray}
Next, we define the meson self-energy function through:
\begin{eqnarray}
\Sigma \left( z,y\right) =-g\underset{h\rightarrow x}{\lim }\left[ 2\partial
_{y}^{\mu }+\partial _{z}^{\mu }\right] \left( \Xi _{3}\right) _{\mu }\left(
z,y,h\right)-g^{2}\eta _{\mu \sigma }\underset{h\rightarrow x}{\lim }\left[
\left( \Xi _{4}\right) ^{\mu \sigma }\left( z,y,h,x\right) -i\delta \left(
y,z\right) \mathscr{D} ^ {\sigma \mu}(h,x)\right] ; \label{eq 2.19}
\end{eqnarray}
with the quantities $\left( \Xi _{3}\right) _{\rho }$ and $\left( \Xi
_{4}\right)  _{\mu \sigma }$ defined by the Eqs.\eqref{eq B.2} and \eqref{eq B.3}, respectively. Hence, by taking the limit of
null sources, having \eqref{eq B.11} as the definition of the scalar propagator, one then find the
following expression to the complete scalar propagator:
\begin{equation}
\mathcal{S}\left( p\right) =\frac{i}{p^{2}-m^{2}-\Sigma \left( p\right) }.
\label{eq 2.20}
\end{equation}

It is easily seen from \eqref{eq 2.20} that the free expression for $\mathcal{S}$ does not differ from the one
obtained on the usual theory. However, the self-energy function \eqref{eq 2.19}, differently of the photon function Eq.\eqref{eq
2.10}, it is sensitive to the effects of the Podolsky $m_{P}$-dependent terms of \eqref{eq 2.15}
already at first-order on perturbation theory.

We also have that the fairly investigated phenomenon in the scalar theory,
the scattering light-by-light, at lowest order, does not change from the
usual theory results once the expressions for $\mathcal{S}$ and 3-points
vertex $\Gamma_{\sigma}$ (as we will show next) are not changed at the tree-level \cite{20}.


\subsection{SDFE for the vertex $\varphi^{\dag}\varphi A$}

\label{sec:3.3}

The starting point for the derivation of the vertex function $\Gamma_{\sigma }$ (defined in \eqref{eq B.12}) is the equation \eqref{eq 2.17},
and its resulting expression also follows from the guideline presented previously. Anyhow,
we should write \eqref{eq 2.17} first in terms of the generating functional $W$, and then after differentiating it with respect to the
source $\bar{\zeta}\left( y\right) $, we finally
take the derivative of the resulting expression with respect to the field $A_{\sigma }\left( z\right) $. However, although the calculation be
straightforward, but quite long, it adds nothing new, neither relevant for the theory discussion.
Hence, we present only its final expression in the following form:
\begin{eqnarray}
\Gamma ^{\sigma }\left( x,y;z\right) =-i\underset{h\rightarrow x}{\lim }%
\left[ 2\partial _{y}^{\sigma }+\partial _{h}^{\sigma }\right] \delta \left(
h,z\right) \delta \left( y,x\right)  +\Lambda ^{\sigma }\left( x,y;z\right) , \label{eq 2.23}
\end{eqnarray}
where we have also defined a new quantity, the vertex part $\Lambda _{\sigma }$ by:
\begin{eqnarray}
\Lambda _{\sigma }\left( x,y;z\right)&=&g\left( \Xi _{6}\right) _{\sigma }\left( x,y,z\right)+\underset{h\rightarrow x}{\lim }%
\left[ 2\partial _{y}^{\mu }+\partial _{h}^{\mu }\right] \left( \Xi
_{5}\right) _{\sigma \mu }\left( x,y,h,z\right) \notag \\
&&-2ig^{2}\delta \left( z,x\right) \int d^{4}sd^{4}f\mathcal{D}^{\alpha
\sigma }\left( f,x\right) \Gamma _{\alpha }\left( s,y;f\right) \mathcal{S}%
\left( x,s\right).   \label{eq 2.24}
\end{eqnarray}%
With the quantities $\left( \Xi _{5}\right) _{\rho }$ and $\left( \Xi
_{6}\right) ^{\mu \sigma }$ defined by the Eqs.\eqref{eq B.4} and \eqref{eq B.5}, respectively.

As it is well known, the SDFE do not only depend on the fundamental Green's functions of a given theory, they do depend on higher-order
functionals, which also satisfies its own SDFE. For instance, the 3-point vertex function depends on the 4-point one, which in
turn depends on the 5-point one, and so on. However, once we are only interested in perturbative calculation, the situation here
is not that complex, since we have only 4 fundamental Green's functions. Also, it is known that all the higher-vertices
can be defined in terms of the fundamental quantities via Feynman diagrams \cite{27}.


\subsection{SDFE for the vertex $\varphi ^{\dag }A\varphi A$}

\label{sec:3.4}

In the same way as for the $3$-point vertex function $\Gamma_{\sigma }$, the starting point to derive $%
\Phi_{\sigma\rho}$ (defined in \eqref{eq B.13}) is to take suitably appropriated functional derivatives of
the equation \eqref{eq 2.17}. Thus, following the same guideline: we write it in
terms of the generating functional $W$, then differentiate it with respect to the source $\bar{\zeta}\left( y\right) $ and we finally
take the derivative of the resulting expression with respect to the fields $A_{\sigma }\left( z\right) $ and $A_{\lambda }\left(
s\right) $. However, the formal development and the evaluation of each term is even longer than as for $\Gamma_{\sigma}$.
Therefore, we present only its final expression casted as:
\begin{eqnarray}%
\Phi _{\sigma \lambda }\left( x,y;z,s\right) =2\eta ^{\sigma \lambda }\delta
\left( s,x\right) \delta \left( z,x\right) \delta \left( x,y\right) \notag +\Psi_{\sigma \lambda }\left( x,y;z,s\right) ,  \label{eq 2.27}
\end{eqnarray}%
with the vertex part $\Psi_{\sigma\lambda}$ defined as the following:
\begin{eqnarray}
\Psi _{\sigma \lambda }\left( x,y;z,s\right) &=&\underset{h\rightarrow x}{%
\lim }\left[ 2\partial _{y}^{\mu }+\partial _{h}^{\mu }\right] \left( \Xi
_{9}\right) _{\mu }^{\lambda \sigma }\left( x,y,z,s,h\right) +\left( \Xi_{10}\right) ^{\lambda \sigma }\left( x,y,z,s\right) \label{eq 2.28} \\
&& -i\delta \left(
x,  y\right) \left( \Xi _{11}\right) ^{\lambda \sigma }\left( z,s,y\right)
+\eta ^{\mu \lambda }\left( \Xi _{7}\right) _{\mu }^{\sigma
}\left( x,y,z\right) \delta \left( s,x\right) +\eta ^{\mu \sigma }\left( \Xi
_{8}\right) _{\mu }^{\lambda }\left( x,s,y\right) \delta \left( z,x\right) . \notag
\end{eqnarray}%
Where the quantities $\left( \Xi _{7}\right) _{\rho }$--$\left( \Xi
_{11}\right) ^{\mu \sigma }$ are defined by the Eqs.\eqref{eq B.6}--\eqref{eq
B.9.0}, respectively.

The conclusions presented previously for $\Gamma_{\sigma}$
also hold to the 4-points vertex function $\Phi_{\sigma\lambda}$.


\section{WARD-FRADKIN-TAKAHASHI IDENTITIES}

\label{sec:4}

The existence of a local gauge symmetry in a field theory generates
constraint relations between the theory's Green's functions. These relations are
known as the Ward-Fradkin-Takahashi identities (WFT). These identities, in terms of the Green's
functions, protect the equivalence on different gauge conditions. Also, as we shall see in the Sec.\ref{sec:5}, these
identities are also strictly related with the renormalizability of a theory.
Hereby, we start the derivation of these identities from the following
identity:
\begin{equation}
\left. \frac{\delta \mathcal{Z}\left[ \zeta ,\bar{\zeta},J_{\mu }\right] }{%
\delta \sigma \left( x\right) }\right\vert _{\sigma =0}=0.  \label{eq 2.34}
\end{equation}%
Thereafter, we find that the generating functional $\mathcal{Z}\left[ \eta ,%
\bar{\eta},J^{\mu }\right] $ satisfies:%
\begin{eqnarray}
\Big[i\frac{\square }{g\xi }\left( 1+m_{P}^{-2}\square \right) ^{2}\partial
_{\mu }\frac{\delta }{\delta J_{\mu }\left( x\right) }+\bar{\zeta}\frac{%
\delta }{\delta \bar{\zeta}\left( x\right) }-\zeta \frac{\delta }{\delta
\zeta \left( x\right) }-\frac{1}{g}\partial _{\mu }J^{\mu }\Big]\mathcal{Z}%
=0.  \label{eq 2.35}
\end{eqnarray}
Next, rewriting \eqref{eq 2.35} as an equation for the generating
functional $W\left[ \zeta ,\bar{\zeta},J_{\mu }\right] $ it follows that:
\begin{eqnarray}
-\frac{\square }{g\xi }\left( 1+m_{P}^{-2}\square \right) ^{2}\partial _{\mu
}\frac{\delta W}{\delta J_{\mu }\left( x\right) }+i\bar{\zeta}\frac{\delta W%
}{\delta \bar{\zeta}\left( x\right) }-i\zeta \frac{\delta W}{\delta \zeta
\left( x\right) }-\frac{1}{g}\partial _{\mu }J^{\mu }=0.  \label{eq 2.36}
\end{eqnarray}
Finally, one can obtain the desired quantum equation of motion for the
theory by writing \eqref{eq 2.36} as an expression for the 1PI-generating functional $\Gamma \left( \varphi ,\varphi ^{\dag },A_{\mu
}\right) $ through the relation \eqref{eq 2.6}. Hence, one gets:
\begin{eqnarray}
-\frac{\square }{g\xi }\left( 1+m_{P}^{-2}\square \right) ^{2}\partial _{\mu
}A^{\mu }\left( x\right) -i\varphi \left( x\right) \frac{\delta \Gamma }{%
\delta \varphi \left( x\right) }+i\varphi ^{\dag }\left( x\right) \frac{%
\delta \Gamma }{\delta \varphi ^{\dag }\left( x\right) }+\frac{1}{g}\partial
_{\mu }\frac{\delta \Gamma }{\delta A_{\mu }\left( x\right) }=0.
\label{eq 2.37}
\end{eqnarray}
In the present theory we actually have three identities, and now we sketch
their derivation. The first identity comes by applying a derivative of $A_{\nu
}\left( y\right) $ in the Eq.\eqref{eq 2.37}:
\begin{equation}
\partial _{\mu }\Gamma ^{\mu \nu }\left( x,y\right) -\frac{\square }{\xi }%
\left( 1+m_{P}^{-2}\square \right) ^{2}\partial ^{\nu }\delta \left(
x,y\right) =0.  \label{eq 2.38}
\end{equation}%
Moreover, the above identity yields to:
\begin{equation}
k_{\mu }\Pi ^{\mu \nu }\left( k\right) =0.  \label{eq 2.40}
\end{equation}%
Equation \eqref{eq 2.12} has been used to obtain the result in the last line. Such result shows the transversal character of the tensor $\Pi ^{\mu
\nu }$. Next, upon the differentiation of \eqref{eq 2.37} with respect
to $\varphi \left( y\right) $ and $\varphi ^{\dag }(z)$, it follows the identity:
\begin{equation}
i\partial _{\mu }\Gamma ^{\mu }\left( z,y;x\right) =\delta \left( x,z\right)
\Gamma \left( x,y\right) -\delta \left( x,y\right) \Gamma \left( x,z\right) ;
\label{eq 2.42}
\end{equation}%
where: $\Gamma \left( x,y\right) =\frac{\delta ^{2}\Gamma }{\delta \varphi
\left( y\right) \delta \varphi ^{\dag }\left( x\right) }$. The last identity to be derived here
follows from the same differentiation as above, but also by differentiating it with respect to $A_{\sigma }\left( w\right) $,
obtaining hence the relation:%
\begin{equation}
i\partial _{\mu }\Phi ^{\mu \sigma }\left( z,y;x,w\right) =\delta \left(
x,z\right) \Gamma ^{\sigma }\left( x,y;w\right) -\delta \left( x,y\right)
\Gamma ^{\sigma }\left( x,z;w\right) .  \label{eq 2.43}
\end{equation}
In possessing of the above identities, Eqs.\eqref{eq 2.42} and \eqref{eq 2.43},
we shall show in the following section how the infinities can be removed from the
$S$-matrix by the renormalization of the fields and physical quantities, such
that the resultant, renormalized $S$-matrix leads to finite values for all
the processes.


\section{RENORMALIZABILITY}

\label{sec:5}

In this section we wish to discuss the overall on-shell renormalization prescription
for the $GSQED_{4}$ \cite{28}. The following analysis will result in state suitable
renormalization conditions, which shall be important to the determination of the renormalization constants in terms of (in)finite
integrals as well.

The bare Lagrangian density is defined in \eqref{eq 1.0}. And introducing
the renormalization constants through the following replacements:
\begin{equation}
\varphi \rightarrow \left( Z_{2}\right) ^{1/2}\varphi ;~~ A\rightarrow
\left( Z_{3}\right) ^{1/2}A,  \label{eq 2.45}
\end{equation}%
we obtain now a fully renormalized Lagrangian:
\begin{eqnarray}
\mathcal{L} &=&\partial _{\mu }\varphi ^{\dag }\partial ^{\mu }\varphi
-m^{2}\varphi ^{\dag }\varphi +ig\varphi ^{\dag }\overleftrightarrow{%
\partial _{\mu }}\varphi A^{\mu } +g^{2}\eta ^{\mu \nu }A_{\mu }\varphi
^{\dag }A_{\nu }\varphi -\frac{1}{4}F_{\mu \nu }F^{\mu \nu }+\frac{1}{%
2m_{P}^{2}}\partial _{\mu }F^{\mu \nu }\partial ^{\sigma }F_{\sigma \nu }\notag \\
&&+\delta _{Z_{2}}\partial _{\mu }\varphi ^{\dag }\partial ^{\mu }\varphi
-\delta _{Z_{0}}m^{2}\varphi ^{\dag }\varphi +i\delta _{Z_{1}}g\varphi
^{\dag }\overleftrightarrow{\partial _{\mu }}\varphi A^{\mu } +\delta _{Z_{4}}g^{2}\eta ^{\mu \nu }A_{\mu }\varphi ^{\dag }A_{\nu }\varphi -\delta
_{Z_{3}}\frac{1}{4}F_{\mu \nu }F^{\mu \nu }.  \label{eq 2.48}
\end{eqnarray}%
Where it was added the counter-terms defined by: $\delta _{Z_{i}}=Z_{i}-1$, it was also introduced the following quantities to the mass:
$Z_{0}m^{2}=Z_{2}m_{0}^{2}$, 3-point vertex: $Z_{1}g=Z_{2}Z_{3}^{1/2}g_{0}$,
and 4-point vertex $Z_{4}g^{2}=Z_{2}Z_{3}g_{0}^{2}$ (associated
with the Compton effect diagrams).\footnote{%
The replacement $\bar{m}_{P}^{2}=Z_{3}m_{P}^{2}$ is only of matter of
notation, once there is not a renormalization constant associated with this parameter.}

The relations between the 3-point and 4-point vertices are compatible if and
only if: $Z_{4}Z_{2}=Z_{1}^{2}$. Furthermore, from the WFT identities, Eqs.%
\eqref{eq 2.42} and \eqref{eq 2.43}, follow the equalities: $%
Z_{1}=Z_{2}$ and $Z_{4}=Z_{1}$, respectively, which are identically satisfied at all order in
perturbation theory. Therefore, the previous relations give at once: $%
Z_{1}=Z_{2}=Z_{4}$. And, thus, we have that the charge renormalization is determined only by: $%
g=Z_{3}^{1/2}g_{0}$.

According to the Lagrangian \eqref{eq 2.48}, one shall obtain new
Schwinger-Dyson-Fradkin equations to the theory. More precisely, the
self-energy functions previously derived are now added by the counter-terms $%
\delta_{Z_{i}}$. We shall denote these new self-energy functions with the
index $^{\left( R\right) }$. Firstly, we will analyze the photon sector,
which the renormalized self-energy function reads:%
\begin{equation}
\Pi ^{\left( R\right) }\left( k\right) =\Pi \left( k\right) +\delta _{Z_{3}},
\label{eq 2.49}
\end{equation}%
where $\Pi \left( k\right) $ is the polarization scalar written in terms of
the renormalized quantities.

The first renormalization condition comes by requiring
that the photon propagator \eqref{eq 2.12}, in the gauge $\xi =1$, must
behave as:%
\begin{equation}
i\mathscr{D}_{\mu \nu }\left( k\right) =\frac{1}{k^{2}}\eta _{\mu \nu
},\quad \text{\ for }k^{2}\rightarrow 0.  \label{eq 2.50}
\end{equation}%
By means of the above requirement one is able to find the expression for the counter-term $\delta_{Z_{3}} $:%
\begin{equation}
\delta _{Z_{3}}=Z_{3}-1=-\left. \Pi \left( k^{2}\right) \right\vert
_{k^{2}\rightarrow 0}.  \label{eq 2.52}
\end{equation}%
It is rather to impose the renormalization conditions of the mesonic sector upon
the 2-point $1PI$-function, which is defined by: $\Gamma \left( p\right)
=p^{2}-m^{2}-\Sigma ^{\left( R\right)}\left( p\right) $. These conditions
are stated as follows:\footnote{$m_{F}$ is defined as the zero of the scalar 2-point $%
1PI$-function.}
\begin{equation}
\frac{\partial \Gamma \left( p\right) }{\partial p^{2}}=1,~~ \Gamma \left( p\right) =p^{2}-m_{F}^{2},\quad \text{when } p^{2}%
\rightarrow m_{F}^{2}.\,  \label{eq 2.55}
\end{equation}
where
\begin{equation}
i\Sigma ^{\left( R\right) }\left( p,m\right) =i\Sigma \left( p,m\right)
-im\delta _{Z_{0}}+i\delta _{Z_{2}}\widehat{p}.
\end{equation}%
Hence, the first condition yields to:\footnote{%
We have used the following decomposition: $\Sigma^{\left( R\right) }\left(
p\right)=\Sigma _{1}^{\left( R\right) }\left( p\right)p^{2}+\Sigma
_{2}^{\left( R\right) }\left( p\right)$.}
\begin{eqnarray}
-\delta _{Z_{2}}=\left. \Sigma _{1}\left( p\right) \right\vert
_{p^{2}\rightarrow m_{F}^{2}}+m_{F}^{2}\left. \frac{\partial \Sigma
_{1}\left( p\right) }{\partial p^{2}}\right\vert _{p^{2}\rightarrow
m_{F}^{2}}+\left. \frac{\partial \Sigma _{2}\left( p\right) }{\partial p^{2}}%
\right\vert _{p^{2}\rightarrow m_{F}^{2}};  \label{eq 2.56}
\end{eqnarray}%
whilst the second one provides the expression:
\begin{eqnarray}
m^{2}\delta _{Z_{0}} =\left. \Sigma _{2}\left(
p\right) \right\vert _{p^{2}\rightarrow m_{F}^{2}}-m_{F}^{2}\Big[
m_{F}^{2}\left. \frac{\partial \Sigma _{1}\left( p\right) }{\partial p^{2}}%
\right\vert _{p^{2}\rightarrow m_{F}^{2}}+\left. \frac{\partial \Sigma _{2}\left( p\right) }{\partial p^{2}}\right\vert _{p^{2}\rightarrow
m_{F}^{2}}\Big] .  \label{eq 2.59}
\end{eqnarray}
Therefore, from the Eqs.\eqref{eq 2.56} and \eqref{eq 2.59} we are able to
compute the renormalization constants $Z_{2}$ and $Z_{0}$, respectively, in
all orders of perturbation theory.

Finally, for the 3-point and 4-point renormalized vertex functions we have
that: at a null transferred momentum limit $k^{2}=\left( p-q\right)
^{2}\rightarrow 0$, the 3-point vertex function is:
\begin{equation}
\Gamma _{\sigma }\left( p,q;k\right) =i\left( p+q\right) _{\sigma },
\label{eq 2.60}
\end{equation}%
whereas for the 4-point vertex function at: $\left( p+k\right) ^{2}=\left(
s+k_{1}\right) ^{2}\rightarrow 0$, it follows that:
\begin{equation}
\Phi _{\sigma \rho }\left( p,s;k,k_{1}\right) =2\eta _{\sigma \rho }.
\label{eq 2.61}
\end{equation}
Thus, with the Eqs.\eqref{eq 2.60} and \eqref{eq 2.61}, one is capable
to obtain the expressions for the renormalization constants $Z_{1}$ and $%
Z_{4}$, respectively. Nevertheless, these constants could already be obtained from the constant
$Z_{2}$, once we have the gauge identity $Z_{1}=Z_{2}=Z_{4}$.

With this section we conclude the formal development of the theory. Henceforth, we will focus
our attention to the explicit evaluation of the radiative correction
expressions. For this purpose, once we are interested on the UV behavior of the
quantities and also in the role played by the higher-order terms on the
self-energy functions, we shall discuss in details the divergent structure of each quantity.

\begin{figure}[tbp]
\begin{center}
\includegraphics[scale=1.0]{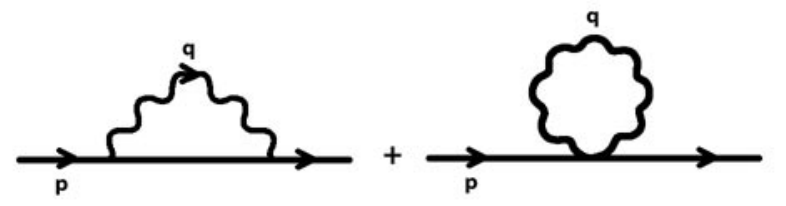}
\end{center}

\caption{Meson self-energy diagrams.}
\label{fig1}
\end{figure}


\section{RADIATIVE CORRECTIONS}

\label{sec:6}

We have so far derived the expressions for the complete Green's functions, the
meson and photon propagators, and the vertex functions for the $GSQED_{4}$, and discussed the renormalizability of these quantities as well.
We are now to consider the evaluation of the radiative correction for these four Green's
functions at $\alpha$-order. As the expression for $\Pi_{\mu\nu}$ at $%
\alpha$-order does not differ from the usual theory, we will present a diagrammatic analysis on the
results at $\alpha^{2}$-order and show, hence, the complete removal of the primitive divergences from the
whole theory. Afterwards, we will discuss whether or not the WFT identities,
Eqs.\eqref{eq 2.42} and \eqref{eq 2.43}, are satisfied, i.e., whether the equalities $Z_{1}=Z_{2}=Z_{4}$ are valid,
once we will observe the presence of a novel and unexpected divergence in the mesonic sector of the theory.

\subsection{Meson self-energy}

\label{sec:6.1}

The first self-energy function that we evaluate here is the one from the mesonic
sector. This quantity corresponds to the diagrams presented in the Fig.\ref{fig1}. In order to
avoid be repetitive in the calculation throughout this section it is
rather interesting to show explicitly some details common for all radiative
functions only in this first discussion.

Anyhow, from the definition \eqref{eq 2.19}, it follows that one can write the self-energy function $\Sigma
$ as $\alpha$-order as:
\begin{eqnarray}
\Sigma ^{\left( 1\right) }\left( p\right) =-ig^{2}\mu ^{\left( 4-\omega
\right) }\int \frac{d^{\omega }q}{\left( 2\pi \right) ^{\omega }}\bigg[
\frac{\left( 2p-q\right) _{\alpha }\left( 2p-q\right) _{\beta }}{\left(
p-q\right) ^{2}-m^{2}+i\epsilon }-2\eta _{\alpha \beta }\bigg] \left[
iD^{\alpha \beta }\left( q\right) \right] .  \label{eq 3.1}
\end{eqnarray}%
We can either combine the terms over the same denominator, and then substituting the
explicit form of the photon propagator $iD^{\alpha \beta }\left( q\right) $, we get the following expression:
\begin{eqnarray}
\Sigma ^{\left( 1\right) }\left( p\right) &=&-ig^{2}\int \frac{d^{4}q}{%
\left( 2\pi \right) ^{4}}\left[ \frac{\left( 2p-q\right) _{\alpha }\left(
2p-q\right) _{\beta }-2\eta _{\alpha \beta }\left(\left( p-q\right)
^{2}-m^{2}\right)}{\left( p-q\right) ^{2}-m^{2}+i\epsilon }\right]  \notag \\
&&\times \left[ \left[ \eta ^{\alpha \beta }-\left( 1-\xi \right) \frac{%
q^{\alpha }q^{\beta }}{q^{2}}\right] \frac{1}{q^{2}}-\left[ \eta ^{\alpha
\beta }-\xi \frac{q^{\alpha }q^{\beta }}{q^{2}-m_{P}^{2}}\right] \frac{1}{%
q^{2}-m_{P}^{2}}+\frac{q^{\alpha }q^{\beta }}{q^{2}\left(
q^{2}-m_{P}^{2}\right) }\right] . \label{eq 3.2}
\end{eqnarray}%
Actually, the first term is inherent from the usual scalar theory (ultraviolet divergent), while the
other two terms are from the Podolsky's theory.\footnote{%
We will always represent the usual contribution through the index $%
\left(1\right)$ in the radiative functions and the new contributions by $%
\left(2\right)$ and $\left(3\right)$.} Once we have already studied the
effects of Podolsky's terms in the fermionic electrodynamics \cite{17}, our
interest now relies on investigating how the second and third terms behave in order to
eliminate or not the ultraviolet divergence from the theory, rendering hence a better UV behavior,
and also whether they give new and interesting information.

Therefore, the above terms are evaluated following the well-known set of
rules of the standard Feynman integrals and dimensional regularization \cite{28}.
Thus, for the first term in \eqref{eq 3.2} we find:\footnote{%
Since we are interested in the divergent behavior of the self-energy
quantities, we would rather to show explicitly only the divergent and constant terms in the text body and presenting
the final expression of the finite terms at the Appendix \ref{sec:C}.}
\begin{eqnarray}
\Sigma ^{\left( 1,1\right) }\left( p\right)&=&\frac{\alpha }{4\pi }\left\{
p^{2}\left[\frac{4}{\epsilon }-2\gamma + \frac{7}{6}\right] +m^{2}\left[
\frac{2}{\epsilon }-\gamma -\frac{7}{2}\right] \right\}  \notag \\
&&-\left( 1-\xi \right) \frac{\alpha }{4\pi }\left\{ p^{2}\left[ -\frac{2}{%
\epsilon }+\gamma +\frac{15}{12}\right] +m^{2}\left[ \frac{2}{\epsilon }-%
\frac{7}{6}-\gamma \right] \right\} +\Sigma _{fin}^{\left( 1,1\right)
}\left( p\right) ;  \label{eq 3.3}
\end{eqnarray}%
with $\alpha=\frac{g^{2}}{4\pi}$, and $\epsilon =4-\omega \rightarrow 0^{+}$ the ultraviolet dimensional regularization
parameter.

Following the same steps as for the previous calculation of $\Sigma
^{\left( 1,1\right) }$, one may obtain the following expressions for the second term:%
\begin{eqnarray}
\Sigma ^{\left( 1,2\right) }\left( p\right) &=&-\frac{\alpha }{4\pi }\bigg\{
p^{2}\left[ \frac{4}{\epsilon }-2\gamma+ \frac{7}{6} \right] +m^{2}\left[
\frac{2}{\epsilon }-\frac{7}{2}-\gamma \right]-\left[ \frac{14}{\epsilon }-%
\frac{1}{2}-7\gamma \right] m_{P}^{2}\bigg\} \notag \\
&& +\xi \frac{\alpha }{4\pi }\bigg\{ p^{2}\left[ -\frac{2}{\epsilon }+\gamma +%
\frac{15}{12}\right]+m^{2}\left[ \frac{2}{\epsilon }-\frac{7}{6}-\gamma %
\right] -\left[ \frac{4}{\epsilon }+\frac{1}{3}-2\gamma \right]
m_{P}^{2}\bigg\} +\Sigma _{fin}^{\left( 1,2\right) }\left( p\right) ,\label{eq 3.4}
\end{eqnarray}%
and for the third one:
\begin{eqnarray}%
\Sigma ^{\left( 1,3\right) }\left( p\right) =\left( 1-2\xi \right) \frac{%
\alpha }{4\pi }\bigg\{ p^{2}\left[ -\frac{2}{\epsilon }+\gamma +\frac{15}{12}%
\right] +m^{2}\left[ \frac{2}{\epsilon }-\frac{7}{6}-\gamma \right] -\left[
\frac{2}{\epsilon }+\frac{1}{6}-\gamma \right] m_{P}^{2}\bigg\} +\Sigma
_{fin}^{\left( 1,3\right) }\left( p\right) .  \label{eq 3.5}
\end{eqnarray}%
Now, by combining the above results, Eqs.\eqref{eq 3.3}, \eqref{eq 3.4} and %
\eqref{eq 3.5} into the equation \eqref{eq 3.2}, it follows that the regularized
expression for the meson self-energy function is given by:
\begin{equation}
\Sigma ^{\left( 1\right) }\left( p\right) =\frac{\alpha }{4\pi }\left[ \frac{%
12}{\epsilon }-\frac{2}{3}-6\gamma \right] m_{P}^{2}+\Sigma _{fin}^{\left(
1\right) }\left( p\right) ,  \label{eq 3.6}
\end{equation}%
where the explicit expression for the finite term: $\Sigma _{fin}^{\left(
1\right) }\left( p\right) =\Sigma_{fin}^{\left( 1,1\right) } \left( p\right)
+\Sigma _{fin}^{\left( 1,2\right)}\left( p\right) +\Sigma _{fin}^{\left(
1,3\right) }\left( p\right) $, is given by the Eq.\eqref{eq C.1}.

Although some contributions from the $m_{P}$-dependent term eliminate the
usual UV divergences (from the $p^{2}$ and $m^{2}$ terms), they also gave
rise to a novel and unexpected gauge-independent logarithmic divergence, which is proportional to the theory's free
parameter $m_{P}$ and therefore it is not present in the conventional theory. Actually, the appearance of this divergence might be
discussed in terms of power counting as well. Nevertheless, this can be a
problematic situation, once we have the equalities between the
renormalization constants to be satisfied. Anyway, we will come back to this problem in the forthcoming Sec.\ref{sec:7}, where it will be
further discussed in details.

\subsection{Photon self-energy at 1-loop}

\label{sec:6.2}

\begin{figure}[tbp]
\begin{center}
\includegraphics[scale=1.1]{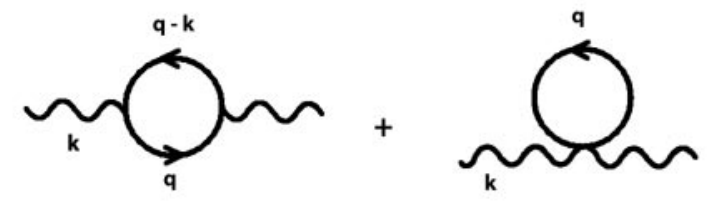}
\end{center}

\caption{Photon self-energy diagrams.}
\label{fig2}
\end{figure}

By means of complementarity, since it has the same expression as the usual
theory, we present here only the final expression of the photon self-energy
function \eqref{eq 2.10} at $\alpha$-order (Fig.\ref{fig2}):
\begin{eqnarray}
\Pi \left( k\right) =-\frac{\alpha }{4\pi }\bigg[ \left[ \frac{2}{\epsilon }%
-\gamma \right] \frac{1}{3} +\int_{0}^{1}dx\left( 1-2x\right) ^{2}\ln \left[
\frac{4\pi \mu ^{2}}{m^{2}-x\left( 1-x\right) k^{2}}\right] \bigg].\label{eq 3.8}
\end{eqnarray}
With $\mu$ being the t'Hooft mass. Regardless its use in the discussion on the running coupling
constant, this result is not relevant for our main interest, once it does not depends on Podolsky's parameter. However, we expect to obtain new and
interesting results for the 2-loop expression of this self-energy function,
once it will possibly depend on this free parameter. This discussion will be presented right below.

\subsection{Vertex part $\Lambda _{\sigma }$}

\label{sec:6.3}

\begin{figure}[tbp]
\begin{center}
\includegraphics[scale=1.075]{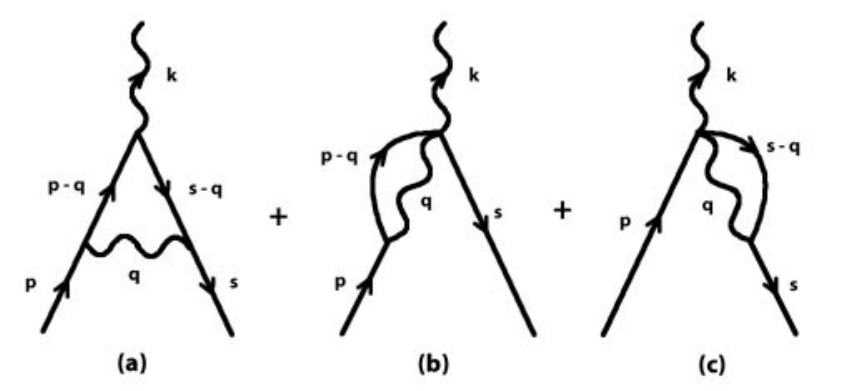}
\end{center}

\caption{Vertex part $\Lambda _{\sigma }$ diagrams.}
\label{fig3}
\end{figure}

We now turn our attention to the calculation of the vertex part $\Lambda _{\sigma }\left( p,s\right)$, Eq.\eqref{eq
2.24}, at $\alpha$-order. Where, as usual, $p$ and $s$ are, respectively, the momenta of the incident and emerging scalar fields,
while $k=p-s$ is the momentum of the photon. We show the corresponding diagrams at Fig.\ref{fig3}.

We have the contribution of three diagrams for the $\Lambda _{\sigma }\left( p,s\right)$ at this order of approximation:
\begin{equation}
\Lambda _{\sigma }\left( p,s\right) =\Lambda _{\sigma }^{\left( a\right)
}\left( p,s\right)+\Lambda _{\sigma }^{\left( b\right) }\left(
p\right)+\Lambda _{\sigma }^{\left( c\right) }\left( s\right) . \label{eq 3.9}
\end{equation}%
But, the diagrams $(b)$ and $(c)$ are related by the momentum symmetry $%
p\leftrightarrow s$. Thereafter, we shall present the calculation and expression for the diagrams $(a)$ and $(b)$ only:
\begin{eqnarray}
\Lambda _{\sigma }^{\left( a\right) }\left( p,s\right)=-ig^{2}\mu ^{\left(
4-\omega \right) }\int \frac{d^{\omega }q}{\left( 2\pi \right) ^{\omega }}%
\bigg[ \frac{\left( 2p-q\right) _{\alpha }\left( 2s-q\right) _{\beta }}{%
\left( s-q\right) ^{2}-m^{2}+i\epsilon }\times \frac{\left( p+s-2q\right) _{\sigma }%
}{\left( p-q\right) ^{2}-m^{2}+i\epsilon }\bigg] \left[ iD^{\alpha \beta
}\left( q\right) \right] ,  \label{eq 3.10}
\end{eqnarray}
and
\begin{eqnarray}
\Lambda _{\sigma }^{\left( b\right) }\left( p\right)=2ig^{2}\mu ^{\left(
4-\omega \right) }\int \frac{d^{\omega }q}{\left( 2\pi \right) ^{\omega }}%
\left[ \frac{\left( 2p-q\right) _{\alpha }\eta _{\beta \sigma }}{\left(
p-q\right) ^{2}-m^{2}+i\epsilon }\right]\left[ iD^{\alpha \beta }\left(
q\right) \right] .  \label{eq 3.11}
\end{eqnarray}
Due to the linear structure of the photon propagator $D_{\sigma\rho}$ the above expressions are decomposed in a sum of three terms,
likewise as happened to the meson self-energy function $\Sigma$, Eq.\eqref{eq 3.2}. Where the first
term is the contribution from the usual Scalar Electrodynamics whilst the second and third ones from the Podolsky's theory.

\subsubsection{$\Lambda _{\sigma }^{\left( a\right) }$}

We calculate first the contribution of the diagram $(a)$, Eq.\eqref{eq 3.10}; where its three terms have the following regularized expressions:
\begin{eqnarray}
\Lambda _{\sigma }^{\left(a,1\right) }\left( p,s\right)=\frac{\alpha }{4\pi
}\bigg[ \frac{2}{\epsilon }-\gamma -\frac{1}{6} -\left( 1-\xi \right) \left(
\frac{1}{\epsilon }-\frac{\gamma }{2}-\frac{5}{12}\right) \bigg]\left(
p+s\right) _{\sigma }+\left( \Lambda _{fin}\right) _{\sigma }^{\left(
a,1\right) }\left( p,s\right),  \label{eq 3.12}
\end{eqnarray}
and
\begin{eqnarray}
\Lambda _{\sigma }^{\left( a,2\right) }\left( p,s\right)=-\frac{\alpha }{%
4\pi }\bigg[ \frac{2}{\epsilon }-\gamma -\frac{1}{6}-\xi \left( \frac{1}{%
\epsilon }-\frac{\gamma }{2}-\frac{5}{12}\right) \bigg] \left( p+s\right)_{\sigma }+\left( \Lambda _{fin}\right) _{\sigma }^{\left( a,2\right)
}\left( p,s\right),  \label{eq 3.13}
\end{eqnarray}
and
\begin{eqnarray}
\Lambda _{\sigma }^{\left( a,3\right) }\left( p,s\right)=\left( 1-2\xi
\right) \frac{\alpha }{4\pi }\left( \frac{1}{\epsilon }-\frac{\gamma }{2}-%
\frac{5}{12}\right) \left( p+s\right) _{\sigma }+\left( \Lambda_{fin}\right) _{\sigma }^{\left( a,3\right) }\left( p,s\right).
\label{eq 3.14}
\end{eqnarray}
Thus, summing up these results, Eqs.\eqref{eq 3.12}, \eqref{eq 3.13} and %
\eqref{eq 3.14}, we determine the first contribution as being finite:
\begin{equation}
\Lambda _{\sigma }^{\left( a\right) }\left( p,s\right)=\left( \Lambda
_{fin}\right) _{\sigma }^{\left( a,1\right) }+\left(
\Lambda _{fin}\right) _{\sigma }^{\left( a,2\right) }+\left( \Lambda _{fin}\right) _{\sigma }^{\left( a,3\right)
},  \label{eq 3.15}
\end{equation}%
where the explicit expression of the quantity $\Lambda ^{\left( a\right) }$ is
given by \eqref{eq C.2}. We see thus that this quantity is already UV finite
and independent of $\mu $ by itself at $\alpha$-order. Now, we expect that
the other two diagrams do present this same finiteness and independence.

\subsubsection{$\Lambda _{\sigma }^{\left( b\right)} $}

Now, for the contribution of the diagram $(b)$, Eq.\eqref{eq 3.11}, it follows the following results:
\begin{equation}
\Lambda _{\sigma }^{\left( b,1\right) }\left( p\right)=-\frac{\alpha }{4\pi }%
p_{\sigma }\left[ \frac{6}{\epsilon }-3\gamma -\frac{1}{3}\left( 1-\xi
\right) \right] +\left( \Lambda _{fin}\right) _{\sigma }^{\left( b,1\right)
}\left( p\right),  \label{eq 3.16}
\end{equation}%
and
\begin{equation}
\Lambda _{\sigma }^{\left( b,2\right) }\left( p\right)=\frac{\alpha }{4\pi }%
p_{\sigma }\left[ \frac{6}{\epsilon }-3\gamma -\frac{1}{3}\xi \right]
+\left( \Lambda _{fin}\right) _{\sigma }^{\left( b,2\right) }\left( p\right),
\label{eq 3.17}
\end{equation}%
and%
\begin{equation}
\Lambda _{\sigma }^{\left( b,3\right) }\left( p\right)=-\left( 1-2\xi
\right) \frac{\alpha }{12\pi }p_{\sigma }+\left( \Lambda _{fin}\right)
_{\sigma }^{\left( b,3\right) }\left( p\right).  \label{eq 3.18}
\end{equation}%
Therefore, by combining the Eqs.\eqref{eq 3.16}, \eqref{eq 3.17} and \eqref{eq
3.18}, one obtains the finite expression:%
\begin{equation}
\Lambda _{\sigma }^{\left( b\right) }\left( p\right)=\left( \Lambda
_{fin}\right) _{\sigma }^{\left( b,1\right) }+\left( \Lambda
_{fin}\right) _{\sigma }^{\left( b,2\right) }+\left( \Lambda
_{fin}\right) _{\sigma }^{\left( b,3\right) },
\label{eq 3.19}
\end{equation}%
with the explicit expression of the quantity $\Lambda ^{\left( b\right) }$ given
by $\left( \ref{eq C.3}\right) $. We see that this quantity is also UV
finite and independent of $\mu $ at $\alpha$-order, as the previous one.

Therefore, through the symmetry property between the diagrams $(b)$ and $%
(c)$, we compute the diagram $(c)$ by taking the limit $p \rightarrow s$ in the equation
\eqref{eq 3.19}. Thus, we have determined that the contributions of these three
diagrams are actually UV finite. However, we have here an apparent violation
of the WFT identities, once we do have the presence of a novel UV divergence in
the resulting expression of $\Sigma$, Eq.\eqref{eq 3.6}, which is not
present, at least, here in the vertex function $\Lambda_{\sigma}$.
Nevertheless, we must now calculate the next vertex function $%
\Phi_{\sigma\rho}$ and evaluate the counter-terms and only
then discuss whether this violation actually do exists or not.

\subsection{ Vertex part $\Psi _{\sigma \rho }$}

\label{sec:6.4}

\begin{figure*}[tbp]
\begin{center}
\includegraphics[scale=1.1]{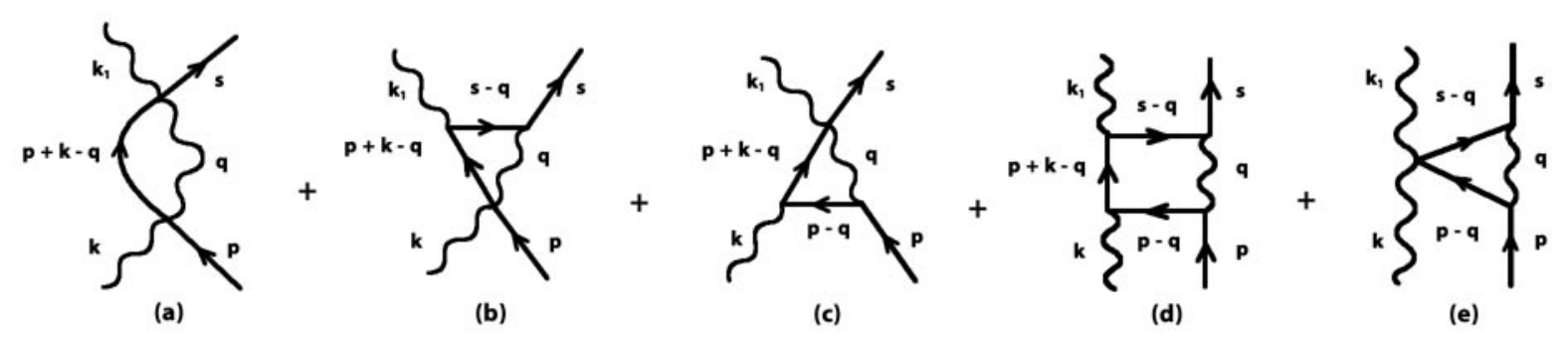}
\end{center}

\caption{Vertex part $\Psi _{\sigma \rho }$ diagrams.}
\label{fig4}
\end{figure*}

The last subject to be discussed in this section, before the analysis upon the
renormalizability of $GSQED_{4}$, is the calculation of the vertex part $%
\Psi_{\sigma\rho}$, Eq.$\left( \ref{eq 2.28}\right) $; which is related to the Compton's scattering. The five diagrams that
contribute to this function at $\alpha$-order are shown in the Fig.\ref{fig4}.

The contributions for the vertex part $\Psi_{\sigma\rho}$ $\left( \ref{eq 2.28}\right) $ at $\alpha$%
-order are given by the sum of terms:
\begin{equation}
\Psi _{\sigma \rho }\left( p,k;s,k_{1}\right) =\Psi _{\sigma \rho }^{\left(
a\right) }+\Psi _{\sigma \rho }^{\left( b\right) }+\Psi _{\sigma \rho
}^{\left( c\right) }+\Psi _{\sigma \rho }^{\left( d\right) }+\Psi _{\sigma \rho }^{\left( e\right) }, \label{eq 3.200}
\end{equation}%
where the explicit expression of each contribution has the following form:%
\begin{eqnarray}
\Psi _{\sigma \rho }^{\left( a\right) }=-4ig^{2}\mu ^{\left( 4-\omega
\right) }\int \frac{d^{\omega }q}{\left( 2\pi \right) ^{\omega }}\left[
\frac{1}{\left( p+k-q\right) ^{2}-m^{2}+i\epsilon }\right] \left[ iD_{\sigma\rho }\left( q\right) \right],  \label{eq 3.20}
\end{eqnarray}
and%
\begin{eqnarray}
\Psi _{\sigma \rho }^{\left( b\right) }=2ig^{2}\mu ^{\left( 4-\omega \right)
}\int \frac{d^{\omega }q}{\left( 2\pi \right) ^{\omega }}\bigg[ \frac{\left(
p+k+s-2q\right) _{\rho }}{\left( s-q\right) ^{2}-m^{2}+i\epsilon }\frac{\left( 2s-q\right) ^{\beta }}{\left( p+k-q\right) ^{2}-m^{2}+i\epsilon }%
\bigg] \left[ iD_{\beta \sigma }\left( q\right) \right] ,  \label{eq 3.21}
\end{eqnarray}
and%
\begin{eqnarray}
\Psi _{\sigma \rho }^{\left( c\right) }=2ig^{2}\mu ^{\left( 4-\omega \right)
}\int \frac{d^{\omega }q}{\left( 2\pi \right) ^{\omega }}\bigg[ \frac{\left(
2p+k-2q\right) _{\sigma }}{\left( p-q\right) ^{2}-m^{2}+i\epsilon }\frac{\left( 2p-q\right) ^{\beta }}{\left( p+k-q\right) ^{2}-m^{2}+i\epsilon }%
\bigg] \left[ iD_{\beta \rho }\left( q\right) \right] ,  \label{eq 3.21a}
\end{eqnarray}
and%
\begin{eqnarray}
\Psi _{\sigma \rho }^{\left( d\right) }=-ig^{2}\mu ^{\left( 4-\omega \right)
}\int \frac{d^{\omega }q}{\left( 2\pi \right) ^{\omega }}\bigg[ \frac{\left(
2p+k-2q\right) _{\sigma }}{\left( p+k-q\right) ^{2}-m^{2}+i\epsilon }\frac{\left( p+k+s-2q\right) _{\rho }}{\left( s-q\right) ^{2}-m^{2}+i\epsilon }%
\frac{\left( 2s-q\right) ^{\beta }\left( 2p-q\right) ^{\alpha }}{\left(
p-q\right) ^{2}-m^{2}+i\epsilon }\bigg] \left[ iD_{\alpha \beta }\left(
q\right) \right] ,\label{eq 3.22}
\end{eqnarray}
at last:%
\begin{eqnarray}
\Psi _{\sigma \rho }^{\left( e\right) }=2i\eta _{\sigma \rho }g^{2}\mu
^{\left( 4-\omega \right) }\int \frac{d^{\omega }q}{\left( 2\pi \right)
^{\omega }}\bigg[ \frac{1}{\left( p-q\right) ^{2}-m^{2}+i\epsilon } \frac{\left( 2p-q\right) _{\alpha }\left( 2s-q\right) _{\beta }}{\left( s-q\right)
^{2}-m^{2}+i\epsilon }\bigg] \left[ iD^{\alpha \beta }\left( q\right) %
\right].  \label{eq 3.23}
\end{eqnarray}
Despite the fact that each one of the contribution of vertex part $
\Psi_{\sigma\rho}$ can be written as the sum of three terms, we shall present directly their final
expressions, as we did for the previous radiative functions. We also have that the contribution $(c)$
can be evaluated through the expression of $(b)$ by taking the limits $s\rightarrow p$ and $\rho \leftrightarrow \sigma$.

\subsubsection{ $\Psi _{\sigma \rho }^{\left(a\right) }$}

The first contribution evaluated is from the diagram $(a)$, Eq.\eqref{eq 3.20}. We have the following regularized result:%
\begin{equation}
\Psi _{\sigma \rho }^{\left( a,1\right) }=\frac{\alpha}{4\pi}\eta _{\sigma \rho }\left[ \left[ \frac{8}{\epsilon }-4\gamma \right]
-\left( 1-\xi \right) \left[ \frac{2}{\epsilon }-\gamma \right] \right]
+\left( \Psi _{fin}\right) _{\sigma \rho }^{\left( a,1\right) },
\label{eq 3.24}
\end{equation}%
and%
\begin{equation}
\Psi _{\sigma \rho }^{\left( a,2\right) }=-\frac{\alpha}{4\pi}\eta _{\sigma \rho }\left[ \left[ \frac{8}{\epsilon }-4\gamma \right]
-\xi \left[ \frac{2}{\epsilon }-\gamma \right] \right] +\left( \Psi
_{fin}\right) _{\sigma \rho }^{\left( a,2\right) },  \label{eq 3.25}
\end{equation}%
and%
\begin{equation}
\Psi _{\sigma \rho }^{\left( a,3\right) }=\left( 1-2\xi \right) \frac{\alpha}{4\pi}\eta _{\sigma \rho }\left[ \frac{2}{\epsilon }%
-\gamma \right] +\left( \Psi _{fin}\right) _{\sigma \rho }^{\left(
a,3\right) }.  \label{eq 3.26}
\end{equation}%
Thus, by combining the Eqs.\eqref{eq 3.24}, \eqref{eq 3.25} and \eqref{eq 3.26}, it follows that:%
\begin{equation}
\Psi _{\sigma \rho }^{\left( a\right) }=\left( \Psi _{fin}\right) _{\sigma
\rho }^{\left( a,1\right) }+\left( \Psi _{fin}\right) _{\sigma \rho
}^{\left( a,2\right) }+\left( \Psi _{fin}\right) _{\sigma \rho }^{\left(
a,3\right) }.  \label{eq 3.27}
\end{equation}%
With the finite term $\Psi ^{\left( a\right) }$ given by the Eq.\eqref{eq C.4}.

\subsubsection{$\Psi _{\sigma\rho }^{\left(b\right) }$ and $\Psi _{\sigma\rho }^{\left(c\right) }$}

Now, evaluating the contribution of the diagram $(b)$, Eq.\eqref{eq 3.21}, it results into:%
\begin{eqnarray}
\Psi _{\sigma \rho }^{\left( b,1\right) }=-\frac{\alpha}{4\pi}\eta _{\sigma \rho }\left[ \left[ \frac{2}{\epsilon }-\gamma \right]
-\left( 1-\xi \right) \left[ \frac{2}{\epsilon }-\gamma -\frac{1}{3}\right] %
\right]+\left( \Psi _{fin}\right) _{\sigma \rho }^{\left( b,1\right) },
\label{eq 3.28}
\end{eqnarray}%
and%
\begin{eqnarray}%
\Psi _{\sigma \rho }^{\left( b,2\right) }=\frac{\alpha}{4\pi}\eta _{\sigma \rho }\left[ \left[ \frac{2}{\epsilon }-\gamma \right]
-\xi \left[ \frac{2}{\epsilon }-\gamma -\frac{1}{3}\right] \right] +\left(\Psi _{fin}\right) _{\sigma \rho }^{\left( b,2\right) },  \label{eq 3.29}
\end{eqnarray}%
and%
\begin{equation}
\Psi _{\sigma \rho }^{\left( b,3\right) }=-\left( 1-2\xi \right) \frac{\alpha}{4\pi}\eta _{\sigma \rho }\left[ \frac{2}{\epsilon }%
-\gamma -\frac{1}{3}\right] +\left( \Psi _{fin}\right) _{\sigma \rho
}^{\left( b,3\right) }.  \label{eq 3.30}
\end{equation}%
We find collecting the Eqs.\eqref{eq 3.28}, \eqref{eq 3.29} and \eqref{eq 3.30}, the finite expression:%
\begin{equation}
\Psi _{\sigma \rho }^{\left( b\right) }=\left( \Psi _{fin}\right) _{\sigma
\rho }^{\left( b,1\right) }+\left( \Psi _{fin}\right) _{\sigma \rho
}^{\left( b,2\right) }+\left( \Psi _{fin}\right) _{\sigma \rho }^{\left(
b,3\right) }.  \label{eq 3.31}
\end{equation}%
Where $\Psi ^{\left( b\right) }$ is given by the Eq.\eqref{eq C.5}. The resulting contribution of $\Psi ^{\left( c\right) }$, Eq.\eqref{eq 3.21a}, is
also given by the finite expression \eqref{eq C.5}, where the appropriated limits: $s\rightarrow p$ and $\rho \leftrightarrow \sigma$ have to be carefully taken.

\subsubsection{ $\Psi _{\sigma \rho }^{\left(d\right) }$}

Next, for the contribution of the diagram $(d)$, Eq.\eqref{eq 3.22}, one obtains:%
\begin{eqnarray}%
\Psi _{\sigma \rho }^{\left( d,1\right) }=\frac{\alpha}{4\pi}\eta _{\sigma \rho }\bigg\{ \left[ \frac{2}{\epsilon }-\gamma %
 -\frac{1}{3}\right]  -\left( 1-\xi \right) \left[  \frac{4}{%
3\epsilon }-\frac{2\gamma }{3} -\frac{5}{9}\right] \bigg\} +\left(
\Psi _{fin}\right) _{\sigma \rho }^{\left( d,1\right) }, \label{eq 3.32}
\end{eqnarray}%
and%
\begin{eqnarray}%
\Psi _{\sigma \rho }^{\left( d,2\right) }=-\frac{\alpha}{4\pi}\eta _{\sigma \rho }\bigg\{ \left[ \frac{2}{\epsilon }-\gamma %
-\frac{1}{3}\right]  -\xi \left[  \frac{4}{3\epsilon }-\frac{%
2\gamma }{3} -\frac{5}{9}\right] \bigg\}+\left( \Psi _{fin}\right)_{\sigma \rho }^{\left( d,2\right) },  \label{eq 3.33}
\end{eqnarray}%
and%
\begin{equation}
\Psi _{\sigma \rho }^{\left( d,3\right) }=\left( 1-2\xi \right) \frac{\alpha}{4\pi}\eta _{\sigma \rho }\left[  \frac{4}{%
3\epsilon }-\frac{2\gamma }{3} -\frac{5}{9}\right] +\left( \Psi
_{fin}\right) _{\sigma \rho }^{\left( d,3\right) }.  \label{eq 3.34}
\end{equation}%
By combining the Eqs.\eqref{eq 3.32}, \eqref{eq 3.33} and \eqref{eq 3.34}, we get the expression:%
\begin{equation}
\Psi _{\sigma \rho }^{\left( d\right) }=\left( \Psi _{fin}\right) _{\sigma
\rho }^{\left( d,1\right) }+\left( \Psi _{fin}\right) _{\sigma \rho
}^{\left( d,2\right) }+\left( \Psi _{fin}\right) _{\sigma \rho }^{\left(
d,3\right) }.  \label{eq 3.35}
\end{equation}%
With the finite term $\Psi ^{\left( d\right) }$ given by the Eq.\eqref{eq C.6}.

\subsubsection{$\Psi _{\sigma \rho }^{\left(e\right) }$}

The last contribution to be evaluated is the corresponding to the diagram $(e)$, Eq.\eqref{eq 3.23}, which yields to:%
\begin{eqnarray}
\Psi _{\sigma \rho }^{\left( e,1\right) }&=&\frac{\alpha}{4\pi}\eta _{\sigma \rho }\left[ \left[ \frac{4}{\epsilon }-1-2\gamma \right]
+\left( 1-\xi \right) \left[ \frac{4}{\epsilon }-\frac{5}{3}-2\gamma \right] %
\right] \notag \\
&&+\left( \Psi _{fin}\right) _{\sigma \rho }^{\left( e,1\right) },
\label{eq 3.36}
\end{eqnarray}
and%
\begin{eqnarray}
\Psi _{\sigma \rho }^{\left( e,2\right) }&=&\frac{\alpha}{4\pi}\eta _{\sigma \rho }\left[ \left[ 1-\frac{4}{\epsilon }+2\gamma \right]
-\xi \left[ \frac{4}{\epsilon }-\frac{5}{3}-2\gamma \right] \right] \notag \\
&&+\left(\Psi _{fin}\right) _{\sigma \rho }^{\left( e,2\right) },  \label{eq 3.37}
\end{eqnarray}
and%
\begin{equation}
\Psi _{\sigma \rho }^{\left( e,3\right) }=-\left( 1-2\xi \right)\frac{\alpha}{4\pi} \eta _{\sigma \rho }\left[ \frac{4}{\epsilon }-%
\frac{5}{3}-2\gamma \right] +\left( \Psi _{fin}\right) _{\sigma \rho
}^{\left( e,3\right) }.  \label{eq 3.38}
\end{equation}%
Finally, summing the Eqs.\eqref{eq 3.36}, \eqref{eq 3.37} and \eqref{eq 3.38},
one obtains:%
\begin{equation}
\Psi _{\sigma \rho }^{\left( e\right) }=\left( \Psi _{fin}\right) _{\sigma
\rho }^{\left( e,1\right) }+\left( \Psi _{fin}\right) _{\sigma \rho
}^{\left( e,2\right) }+\left( \Psi _{fin}\right) _{\sigma \rho }^{\left(
e,3\right) }.  \label{eq 3.39}
\end{equation}%
Where the finite term $\Psi ^{\left( e\right) }$ is given by Eq.\eqref{eq
C.7}.

From the above results for the radiative correction expressions at $\alpha$-order, we see that the primitive divergent self-energy diagrams
are only from the meson and photon self-energy functions, Eqs.\eqref{eq 3.6} and %
\eqref{eq 3.8}, respectively. As a matter of fact, to elucidate this last point, we shall present next a
discussion upon the photon self-energy at $\alpha^{2}$-order, which has contributions from the meson, and 3-point and 4-point vertex functions.
However, we will find another primitive divergence on the photon self-energy, but now $m_{P}$-dependent.


\subsection{Photon self-energy at 2-loop}
\label{sec:6.5}

\begin{figure*}[tbp]
\begin{center}
\includegraphics[scale=1]{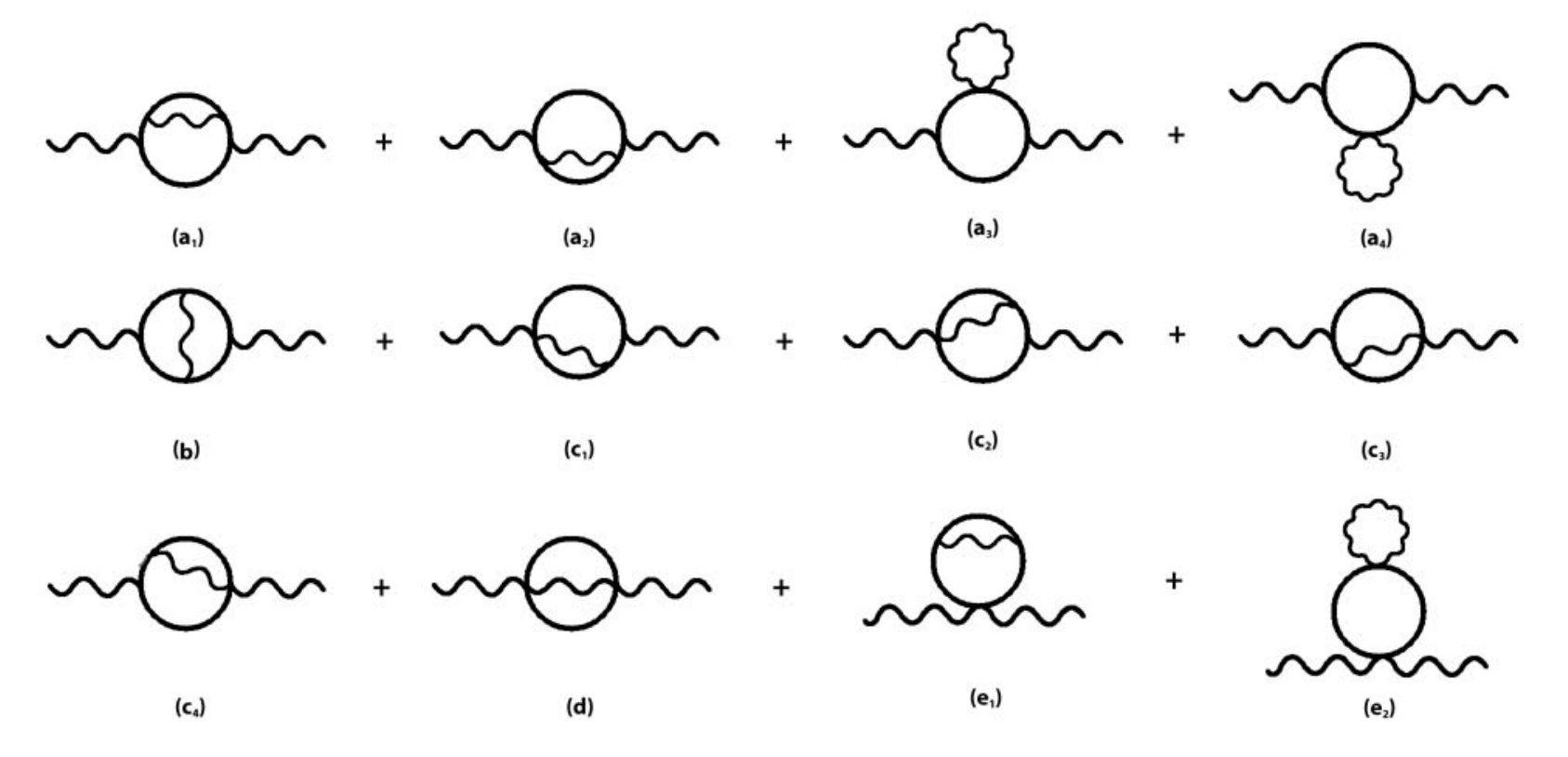}
\end{center}

\caption{Photon self-energy diagrams at $\alpha^{2}$-order.}
\label{fig5}
\end{figure*}

\begin{figure*}[tbp]
\begin{center}
\includegraphics[scale=1]{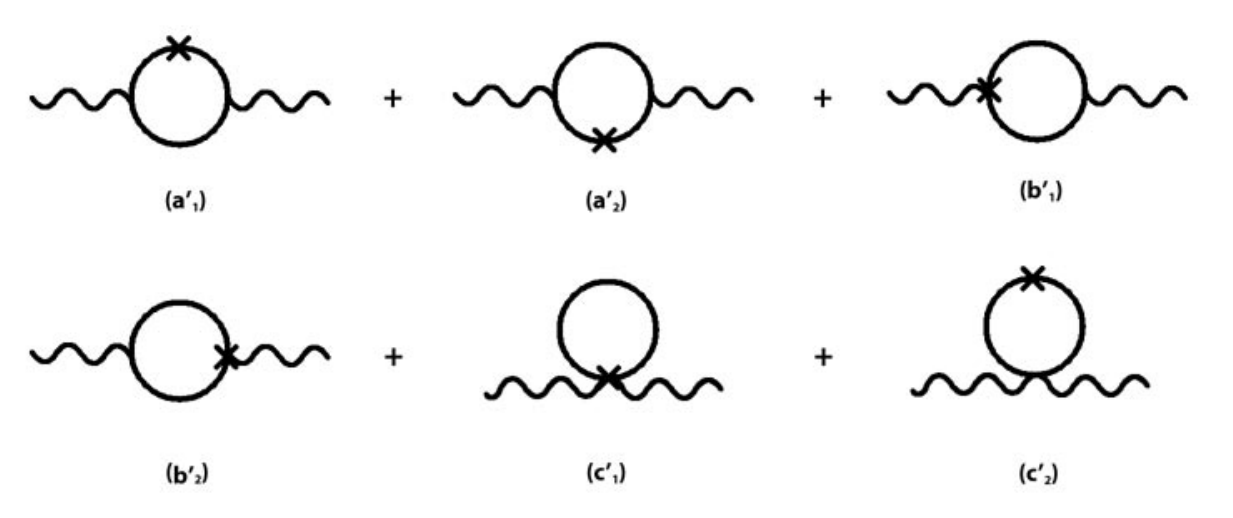}
\end{center}

\caption{Photon counter-terms diagrams at $\alpha^{2}$-order.}
\label{fig6}
\end{figure*}

In comparison with the usual theory and also with $GQED_{4}$, the $GSQED_{4}$ presents an unexpected novel divergent structure.
In this sense, it is important to supply the discussion of this new kind of divergence ($m_{P}$-dependent one) present in the electron
self-energy Eq.\eqref{eq 3.6}, with new information and details. Therefore, for this purpose, we will present a diagrammatic analysis on the
results of $\alpha^{2}$-order photon self-energy. Our interest in this particular function is driven mainly by the divergence structure
embedded on it from the meson self-function (the set of diagrams $(a)$ and $(e)$). Furthermore, we are motivated to present also a discussion
on the behavior of this self-energy function in face of the higher-order terms, once the $\alpha$-order calculation is not sensitive to these effects.

The diagrams presented in this order are depicted at the Fig.\ref{fig5} and the counter-term diagrams at the Fig.\ref{fig6}. It is worth
to stress that we do not have the intention to present here a formal proof concerning the complete renormalizability of the theory. Nevertheless,
we believe that a qualitative discussion does provide all the necessary information concerning the renormalizability of the theory, especially
regarding the $m_{P}$-dependent divergent diagrams.

First, as it is easily seen, we do have that the set of diagrams $(a)$ and $(e)$ are contributions which come from the meson self-energy $\Sigma$. It
is not difficult to show through their divergent structure that the diagrams $(a)$ and $(e)$ are UV divergent, being $m_{P}$-dependent
(here the other divergences are also canceled as the previous self-energy functions) and, as it turns out, it is absorbed by
the counter-terms diagrams $(a_{1}^{\prime})$, $(a_{2}^{\prime})$ and $(c_{2}^{\prime})$.

Next, we do have that the set of diagrams $(b)$ and $(c)$ carry contribution from the $3$-point vertex $\Lambda$. Thus, after a calculation one
obtains that the diagrams $(b)$ and $(c)$ contain a $m_{P}$-dependent UV divergence. Divergence which is absorbed by the
counter-terms diagrams $(b_{1}^{\prime})$ and $(b_{2}^{\prime})$.

The remaining contribution to be analyzed is from the diagram $(d)$. Actually, the contribution $(d)$ is also a primitive divergence of the photon
self-energy diagram, and the $m_{P}$-dependent ultraviolet divergence presented in this diagram is absorbed by
the counter-term diagram $(c_{1}^{\prime})$. With the above results, we see that no further counter-terms are needed, therefore, the
renormalizability of the theory is consistently stated.


\section{RENORMALIZED QUANTITIES}

\label{sec:7}

In this section we finally present the explicit expressions and discussion on some renormalized quantities. We start by presenting the
expressions for the counter-terms evaluated at $\alpha$-order. Also, from the counter-terms
expressions we will be able to recognize that the novel logarithmic divergence present at the meson self-energy, Eq.\eqref{eq 3.6}, actually does not spoil
the WFT identities, once it is removed by the mass counter-term $\delta_{Z_{0}}$. Afterwards, we discuss the effective coupling of the $GSQED_{4}$ and
its relation with the $GQED_{4}$ as well. Furthermore, from the running coupling constant expression we will be able to determine a energy range
where the theory is well-defined.

\subsection{ Counter-terms expressions}

\label{sec:7.}

\subsubsection{ $\delta _{Z_{3}}$}

 \label{sec:7.0}

First, we have the simplest counter-term expression at $\alpha$-order, which is computed
from the condition \eqref{eq 2.52} and result \eqref{eq 3.8}; in this way it follows:
\begin{equation}
\left( Z_{3}-1\right) ^{\left( 1\right) }=\frac{\alpha }{12\pi }\left[ \frac{%
2}{\epsilon }-\gamma +\ln \left[ \frac{4\pi \mu ^{2}}{m^{2}}\right] \right] ;
\label{eq 4.0}
\end{equation}
showing that the ultraviolet divergence of the photon propagator is absorbed properly by its counter-term.

\subsubsection{$\delta _{Z_{0}}$}

 \label{sec:7.1}

Next, we compute the expression for the counter-term $\delta
_{Z_{0}}$ through the condition \eqref{eq
2.59} and expression \eqref{eq 3.6}, and it follows for $b=\frac{m_{P}^{2}}{m^{2}}>4$ the expression:
\begin{eqnarray}
\delta _{Z_{0}}^{\left( 1\right) }&=&\frac{\alpha }{2\pi } \frac{\left( \xi
-3\right)}{\epsilon_{IR}}+\frac{\alpha }{4\pi }\left\{ \frac{12}{%
\epsilon }-\frac{2}{3}-6\gamma +13\ln \left[
\frac{4\pi \mu ^{2}}{m^{2}}\right]\right\} \left( \frac{m_{P}^{2}}{m^{2}}\right)\notag \\
&& +\frac{\alpha }{24\pi }\left\{ 144b^{3}\xi +126b^{2}\left( 1-4\xi \right)
+2b\left( 99\xi +14\right)  -6\right\}   \nonumber \\
&&+\frac{\alpha }{4\pi }\left( -12b^{4}\xi +\frac{3}{2}b^{3}(40\xi
-7)+b^{2}(20-66\xi )+b\left( 6\xi -\frac{3}{2}\right) +\xi -3\right) \log (b)\notag\\
&&+\frac{\alpha }{8\pi }\frac{b}{\sqrt{b\left( b-4\right) }}\left(
24b^{4}\xi -21b^{3}(8\xi -1)+b^{2}(324\xi -82)-3b(44\xi -5)+20\right)\notag \\
&&\times \left[ \log \left( \frac{b-2+\sqrt{(b-4)b}}{b-2-\sqrt{(b-4)b}}%
\right) +\log \left( \frac{b-\sqrt{(b-4)b}}{b+\sqrt{(b-4)b}}\right) \right] %
.  \label{eq 4.2}
\end{eqnarray}
Otherwise, the logarithm must be replaced by $arctan$ function. We have introduced the infrared dimensional parameter as
$\epsilon_{IR}=\omega-4$, $\epsilon_{IR} \rightarrow 0^{-}$. Finally, as it turns out, we see explicitly in the above expression,
Eq.\eqref{eq 4.2}, that the novel $m_{P}$-dependent logarithmic divergence of the meson propagator is actually absorbed by the counter-term $\delta _{Z_{0}}$.
Therefore this divergence poses no harm to the theory's renormalizability, once the gauge WFT identities are still satisfied.

\subsubsection{ $\delta _{Z_{2}}$}

\label{sec:7.2}

At last, we compute $\delta _{Z_{2}}$ through \eqref{eq 2.56} and \eqref{eq 3.6}. Thus, under the condition $b=\frac{m_{P}^{2}}{%
m^{2}}>4$ it follows that:
\begin{eqnarray}
\delta _{Z_{2}}^{\left( 1\right) }&=&\frac{\alpha }{2\pi }\frac{\left( \xi
-3\right)}{\epsilon_{IR}}+\frac{\alpha }{2400\pi }\bigg\{720b^{4}\left( 2\xi -1\right)
+60b^{3}\left( 81-202\xi \right) +30b^{2}\left( 858\xi -19\right) \notag\\
&&-20b\left( 9+577\xi \right) +741-1482\xi \bigg\}  -\frac{\alpha }{80\pi }\bigg\{ 12b^{5}\left( 2\xi -1\right) +b^{4}\left(
99-238\xi \right) \nonumber \\
&& +2b^{3}\left( 353\xi -59\right) -b^{2}\left( 668\xi
+51\right) +10b\left( 13\xi -6\right) +20\left(3-\xi \right)\bigg\} \log \left[ b\right]
\notag \\
&&+\frac{\alpha }{80\pi }\frac{b}{\sqrt{b\left( b-4\right) }}\bigg\{
12b^{5}\left( 2\xi -1\right) +b^{4}\left( 123-286\xi \right) +2b^{3}\left(
567\xi -146\right) +5b^{2}\left( 7-340\xi \right)   \notag \\
&&+2b\left( 1+383\xi \right)
+40\left( 1-\xi \right) \bigg\}  \left[ \log \left[ \frac{b-\sqrt{b\left( b-4\right) }}{b+\sqrt{%
b\left( b-4\right) }}\right] +\log \left[ \frac{b+\sqrt{(b-4)b}-2}{b-\sqrt{%
(b-4)b}-2}\right] \right]  . \label{eq 4.1}
\end{eqnarray}%
Such expression is also an ultraviolet finite quantity at order-$\alpha$.

Actually, by computing the counter-terms $\delta _{Z_{1}}$ and $\delta _{Z_{4}}$ through the conditions \eqref{eq 2.60} and \eqref{eq 2.61}, respectively, we obtain, for both
quantities, the same expression that \eqref{eq 4.1} at $\alpha$-order. Therefore, such results are in agreement with the gauge WFT identities,
and despite the fact that the meson propagator presents an unexpected divergence, it was shown that it is suitably absorbed by its
proper counter-term. We also observe that the above counter-terms expressions, \eqref{eq 4.2} and \eqref{eq 4.1}, are also infrared
finite at the Fried-Yennie gauge choice: $\xi=3$ \cite{29}. Therefore, with these results we see that the renormalizability
of $GSQED_{4}$ is attained (even to the $2$-loop photon self-energy results) and the divergences are consistently absorbed by the appropriated counter-terms.

\subsection{ Effective coupling constant}
\label{sec:7.4}

Although the renormalization constant $Z_{3}$ expression at $\alpha$-order, Eq.\eqref{eq 4.0},
does not feel the effects from the higher-order terms, there are other quantities that may present modifications from the usual
expression at this order. One of these quantities is the Born amplitude \cite{19}.
Actually, this renormalized amplitude depends on $Z_{3}$, which provides a suitable context
for our discussion. Also, from this analysis we will be able to introduce an
invariant quantity, the so-called running coupling constant for the $GSQED_{4}$; and from the later, we shall determine an energy range for the theory.

By means of simplicity, we will work in the Landau gauge $\xi=0$.
Therefore, from the photon propagator, in the Born scattering and $k^{2}\gg m^{2}$ regime, one can define in this approximation the running coupling constant as:
\begin{eqnarray}
\alpha _{R}\left( k^{2}\right) = \alpha \bigg[ 1+\left( 1-\frac{1}{k^{2}-m_{P}^{2}}\right)
\bigg[ -\frac{\alpha }{12\pi }\bigg[ \frac{2}{\epsilon }-\gamma +\ln \left[\frac{4\pi \mu ^{2}}{m^{2}}\right] \bigg] +\frac{\alpha }{12\pi }\ln \left(
\frac{k^{2}}{m^{2}}\right) \bigg] \bigg] , \notag
\end{eqnarray}%
for which can immediately be casted as:
\begin{equation}
\alpha _{R}\left( k^{2}\right) =\alpha _{R}\left( m^{2}\right) \left[ 1+%
\frac{\alpha _{R}\left( m^{2}\right) }{12\pi }\frac{1}{1-\frac{k^{2}}{%
m_{P}^{2}}}\ln \left[ \frac{k^{2}}{m^{2}}\right] \right] ,  \label{eq 4.5}
\end{equation}
where $\alpha _{R}\left( m^{2}\right)=Z_{3}\alpha$, with $Z_{3}$ given by the Eq.\eqref{eq 4.0}.

Nevertheless, further modifications to the Born scattering can also be
studied in subsequent orders of perturbation theory. Thus, in order to obtain
these higher-orders modifications, we can sum an important class of
diagrams, which consists in the most divergent set of logarithms. Therefore, the
running coupling constant expression, in such regime and approximation, can
be casted in the following form:
\begin{equation}
\frac{1}{\alpha _{R}\left( k^{2}\right) }=\frac{1}{\alpha _{R}\left(
m^{2}\right) }-\frac{1}{12\pi }\frac{1}{1-\frac{k^{2}}{m_{P}^{2}}}\ln \left[
\frac{k^{2}}{m^{2}}\right] .  \label{eq 4.7}
\end{equation}

Exactly as it happens in the usual scalar theory in relation to the $QED$, the rate of change of the coupling constant in $GSQED_{4}$
is one-fourth of that from $GQED_{4}$ \cite{18}. Also, we see the presence of a pole at $m_{P}^{2} = k^{2} $ on its expression; and, in
comparison to the $SQED_{4}$ expression \cite{19} it provides a validity regime: $m^{2} \leq k^{2} < m_{P}^{2}$, where the generalized
theory is in fact well-defined.

\section{CONCLUDING REMARKS}

\label{sec:8}

Despite the spinless aspect of the Generalized Scalar Electrodynamics, from our present analysis it has provided insights of a particular HD
term in an interesting field theory, when the $GSQED_{4}$ is discussed in the context of effective theories.
Foremost is the role played by the HD and the generalized gauge condition as well, that showed significant importance when the radiative
corrections were computed at $\alpha$-order. Also, the presence of a new type of divergence in the mesonic sector is a result of some interest in itself.

In this paper, we presented a proper study regarding the complete quantization and consistent renormalization, and subsequent consequences,
of the Generalized Scalar Electrodynamics. The first part of this article was devoted to the formal development of $GSQED_{4}$. After we have
constructed the transition-amplitude through the Batalin-Fradkin-Vilkovisky covariant method, we derived by functional methods the four
fundamental Green's functions for the theory and the Ward-Fradkin-Takahashi identities as well. And, our main motivation here relied
on observing the theory's behavior upon these $m_{P}$-dependent terms, once the scalar theory possesses a more interesting interacting
structure. It was also shown that the two gauge WFT identities, Eqs.\eqref{eq 2.42} and \eqref{eq 2.43}, were of highly importance on the
discussion regarding the renormalizability of $GSQED_{4}$. On the discussion about the renormalizability of $GSQED_{4}$ we made use
of a previous result on the renormalization condition for the photon propagator \cite{18}.

Well, in the second part of the article it was the calculation of radiative corrections that took place, we presented there the evaluation and
discussion of the results of the self-energy functions and counter-terms at $\alpha$-order. We initially discussed the results on the
meson self-energy, where we found that the massive terms canceled the usual ultraviolet divergences but also generated a novel
logarithmic divergence, which is proportional to the free parameter $m_{P}$ and therefore not present in the conventional theory. However, for the vertex parts $\Lambda$ and $\Psi$ it was found that the massive contribution of the photon propagator canceled out all the divergences, resulting, thus, in an ultraviolet finite expression for both quantities. Moreover, once the massive contributions were not presented in the photon self-energy at $\alpha$-order, we provided
a diagrammatic discussion about the contribution of the diagrams in $\alpha^{2}$-order and its counter-terms, where it is also found $m_{P}$-dependent divergence;
but all of them are properly absorbed by the theory's counter-terms. Next, we presented the explicit expressions for the counter-terms at
$\alpha$-order. Actually, we have shown that the $m_{P}$-dependent divergence from the meson self-energy function was absorbed suitably by
$\delta_{Z_{0}}$; showing, therefore, that the gauge
WFT identities ($Z_{1}$=$Z_{2}$=$Z_{4}$) are still satisfied at this order. Also, the expressions for the counter-terms: $\delta _{Z_{0}}$,
$\delta _{Z_{2}}$, $\delta_{Z_{1}}$ and  $\delta_{Z_{4}}$, are infrared finite at the Fried-Yennie gauge, $\xi =3$. At last, from the
effective coupling expression we have found a energy scale: $m^{2} \leq k^{2} < m_{P}^{2}$, where the generalized theory is in fact well-defined.

Once again, the Podolsky's theory has shown to possess a richness of features, and be interesting in its own right. Here we have
successfully studied the generalized electrodynamics of spinless charged particles in rich details. And in possess of such results we do have now good insights
to deal also with non-abelian fields, for instance, since the scalar theory shares some general properties with the non-abelian one \cite{31}.
However, this study might take place not only in a four-dimensional space-time, but also in different dimensionality, once the field theories
in lower dimensions are again receiving attention nowadays. Currently, based on the present results, the authors are also investigating
the Scalar Generalized Electrodynamics also at a thermodynamical equilibrium. An interesting and still unexplored issue that we think that
deserves attention next, is the study of scattering processes with external fields of the Podolsky's theory, either in the context
of spinor and scalar electrodynamics \cite{32}. These issues and others will be further elaborated, investigated and reported elsewhere.


\subsection*{ACKNOWLEDGEMENTS}

The authors thank to an anonymous referee for the comments and suggestions, and
Daniel Soto Barrientos for the fruitful discussions. RB thanks FAPESP for full support and BMP
thanks CNPq and CAPES for partial support.


\appendix


\section{Functional Quantities}

\label{sec:B}

In the definition of the polarization tensor $\Pi $, Eq.\eqref{eq 2.10}, we have introduced by convenience the following quantities:{\small
\begin{eqnarray}
\left( \Xi _{1}\right) _{\rho }\left( x,h;z\right) =g\int d^{4}wd^{4}r\mathcal{S}\left( r,h\right) \Gamma _{\rho }\left(
w,r;z\right) \mathcal{S}\left( x,w\right) ,  \label{eq B.0}
\end{eqnarray}
}and{\small
\begin{eqnarray}
\left( \Xi _{2}\right) _{\lambda }^{\mu }\left( x,h;z\right) &=&-g^{2}\int d^{4}wd^{4}vd^{4}rd^{4}g_{1}d^{4}u \bigg\{\mathcal{S}\left(
g_{1},h\right) \mathcal{D}^{\mu \rho }\left( x,u\right) \Gamma _{\rho
}\left( r,g_{1};u\right) \mathcal{S}\left( v,r\right) \Gamma _{\lambda
}\left( w,v;z\right) \mathcal{S}\left( x,w\right)  \notag \\
&&+\mathcal{S}\left( v,h\right) \Gamma _{\lambda }\left( w,v;z\right)
\mathcal{S}\left( g_{1},w\right) \mathcal{D}^{\mu \rho }\left( x,u\right)
\Gamma _{\rho }\left( r,g_{1};u\right) \mathcal{S}\left( x,r\right) \bigg\}\notag \\
&&+ig^{2}\int d^{4}wd^{4}vd^{4}u\mathcal{S}\left( v,h\right) \mathcal{D}%
_{\rho }^{\mu }\left( x,u\right) \Phi _{\lambda \rho }\left( w,x;z,u\right)
\mathcal{S}\left( x,w\right) \label{eq B.1}.
\end{eqnarray}%
}Also, in the definition of meson self-energy $\Sigma $, Eq.\eqref{eq 2.19}, we have introduced:{\small
\begin{equation}
\left( \Xi _{3}\right) _{\mu }\left( z,y,h\right) =g\int
d^{4}sd^{4}f\mathcal{S}\left( s,y\right) \mathcal{D}_{\alpha \mu }\left(
f,h\right) \Gamma ^{\alpha }\left( z,s;f\right) ,  \label{eq B.2}
\end{equation}%
}and{\small
\begin{eqnarray}
\left( \Xi _{4}\right) ^{\mu \sigma }\left( z,y,h,x\right) &=& -g^{2}\int d^{4}rd^{4}vd^{4}fd^{4}cd^{4}u \bigg\{\mathcal{S}\left( c,y\right)
\mathcal{D}^{\rho \mu }\left( u,x\right) \Gamma _{\rho }\left( r,g;u\right)
\mathcal{S}\left( v,r\right) \mathcal{D}^{\alpha \sigma }\left( f,h\right)
\Gamma _{\alpha }\left( z,v;f\right) \notag \\
&&+\mathcal{S}\left( v,y\right) \mathcal{D}^{\alpha \sigma }\left(
f,h\right) \Gamma _{\alpha }\left( r,v;f\right) \mathcal{S}\left( c,r\right)
\mathcal{D}^{\rho \mu }\left( u,x\right) \Gamma _{\rho }\left( z,g;u\right) %
\bigg\}\notag \\
&&+ig^{2}\int d^{4}vd^{4}fd^{4}u\mathcal{S}\left( v,y\right) \mathcal{D}%
^{\alpha \sigma }\left( f,h\right) \mathcal{D}^{\rho \mu }\left( u,x\right)
\Phi _{\alpha \rho }\left( z,v;f,u\right) . \label{eq B.3}
\end{eqnarray}%
}Furthermore, in the definition of vertex functions $\Lambda $ and $\Psi $, Eqs.\eqref{eq 2.24} and \eqref{eq 2.28}, respectively, we have introduced the
quantities: $\Xi _{5}-\Xi _{11}$, which are actually lengthy, and their derivation and content do not add nothing to the formal development of the theory. Also, all of
them can be computed by following the standard guideline of functional methods presented here. Thus, we have their definition read as:{\small
\begin{eqnarray}%
\left( \Xi _{5}\right) _{\mu }^{\sigma }\left( x,y,h,z\right) = \int d^{4}w%
\frac{\delta ^{2}\Gamma }{\delta \varphi \left( y\right) \delta \varphi
^{\dag }\left( w\right) } \frac{\delta }{\delta J^{\mu }\left( h\right) }%
\left[ \frac{\delta }{\delta A_{\sigma }\left( z\right) }\left( \frac{\delta
^{2}W}{\delta \zeta \left( w\right) \delta \bar{\zeta}\left( x\right) }%
\right) \right] ,  \label{eq B.4}
\end{eqnarray}%
}and{\small
\begin{eqnarray}%
\left( \Xi _{6}\right) _{\mu \nu }^{\sigma }\left( x,y,z\right) =\underset{t\rightarrow x}{\underset{h\rightarrow x}{\lim }}\int d^{4}w\frac{%
\delta ^{2}\Gamma }{\delta \varphi \left( y\right) \delta \varphi ^{\dag
}\left( w\right) }\frac{\delta ^{2}}{\delta J^{\nu }\left( t\right) \delta
J^{\mu }\left( h\right) }\left[ \frac{\delta }{\delta A_{\sigma }\left(
z\right) }\left( \frac{\delta ^{2}W}{\delta \zeta \left( w\right) \delta
\bar{\zeta}\left( x\right) }\right) \right] , \label{eq B.5}
\end{eqnarray}%
}and{\small
\begin{eqnarray}%
\left( \Xi _{7}\right) _{\mu }^{\sigma }\left( x,y,z\right) =2i\int d^{4}w%
\frac{\delta ^{2}\Gamma }{\delta \varphi \left( y\right) \delta \varphi
^{\dag }\left( w\right) }  \frac{\delta }{\delta A_{\sigma }\left( z\right) }%
\left( \frac{\delta ^{3}W}{\delta J^{\mu }\left( x\right) \delta \zeta
\left( w\right) \delta \bar{\zeta}\left( x\right) }\right) , \label{eq B.6}
\end{eqnarray}%
}and{\small
\begin{eqnarray}%
\left( \Xi _{8}\right) _{\mu }^{\lambda }\left( x,y,s\right) ={\underset{h\rightarrow x}{\lim }} \eta^{\lambda}_{\sigma}
\left( \Xi _{7}\right) _{\mu }^{\sigma }\left( x,y,h\right), \label{eq B.7}
\end{eqnarray}%
}and{\small
\begin{eqnarray}%
g\left( \Xi _{9}\right) _{\mu }^{\lambda \sigma }\left( x,y,z,s,h\right)
=\int d^{4}w\frac{\delta ^{2}\Gamma }{\delta \varphi \left( y\right) \delta
\varphi ^{\dag }\left( w\right) }  \frac{\delta }{\delta J^{\mu }\left(
h\right) }\left[ \frac{\delta }{\delta A_{\lambda }\left( s\right) }\left[
\frac{\delta }{\delta A_{\sigma }\left( z\right) }\left( \frac{\delta ^{2}W}{%
\delta \zeta \left( w\right) \delta \bar{\zeta}\left( x\right) }\right) %
\right] \right] , \label{eq B.8}
\end{eqnarray}%
}and{\small
\begin{eqnarray}%
\left( \Xi _{10}\right) ^{\lambda \sigma }\left( x,y,z,s\right) =\underset{h\rightarrow x}{\underset{f\rightarrow x}{\lim }}\int d^{4}w\frac{%
\delta ^{2}\Gamma }{\delta \varphi \left( y\right) \delta \varphi ^{\dag
}\left( w\right) } \frac{\delta }{\delta J_{\mu }\left( f\right) }\left[
\frac{\delta }{\delta J^{\mu }\left( h\right) }\left[ \frac{\delta }{\delta
A_{\lambda }\left( s\right) }\left[ \frac{\delta }{\delta A_{\sigma }\left(
z\right) }\left( \frac{\delta ^{2}W}{\delta \zeta \left( w\right) \delta
\bar{\zeta}\left( x\right) }\right) \right] \right] \right],  \label{eq B.9}
\end{eqnarray}%
}at last:{\small
\begin{eqnarray}%
 \left( \Xi _{11}\right) ^{\lambda \sigma }\left( y,z,s\right) =\underset{h\rightarrow x}{\underset{f\rightarrow x}{\lim }}\int d^{4}w\frac{%
\delta ^{2}\Gamma }{\delta \varphi \left( y\right) \delta \varphi ^{\dag
}\left( w\right) } \frac{\delta }{\delta A_{\lambda }\left( s\right) }\left[
\frac{\delta }{\delta A_{\sigma }\left( z\right) }\left[ \frac{\delta ^{2}W}{%
\delta J_{\mu }\left( f\right) \delta J^{\mu }\left( h\right) }\right] %
\right] . \label{eq B.9.0}
\end{eqnarray}%
}Also, within the formal development, and in the above quantities as well, it has been used the following definitions for the connected two-point functions:{\small
\begin{equation}
i\mathcal{D}_{\sigma \mu }\left( x,y\right) =\left. \frac{\delta ^{2}W}{%
\delta J_{\mu }\left( y\right) \delta J_{\sigma }\left( x\right) }%
\right\vert _{s=0} ,  \label{eq B.10}
\end{equation}%
}and{\small
\begin{equation}
i\mathcal{S}\left( x,y\right) =\left. \frac{\delta ^{2}W}{\delta \zeta
\left( y\right) \delta \bar{\zeta}\left( x\right) }\right\vert _{s=0},
\label{eq B.11}
\end{equation}%
}and for the $1PI$ vertex functions:{\small
\begin{equation}
g\Gamma _{\sigma }\left( x,y;z\right) =\left. \frac{\delta ^{3}\Gamma }{%
\delta A_{\sigma }\left( z\right) \delta \varphi ^{\dag }\left( y\right)
\delta \varphi \left( x\right) }\right\vert _{f=0},  \label{eq B.12}
\end{equation}
}and{\small
\begin{equation}
g^{2}\Phi _{\sigma \rho }\left( x,y;z,w\right) =\left. \frac{\delta
^{4}\Gamma }{\delta A_{\rho }\left( w\right) \delta A_{\sigma }\left(
z\right) \delta \varphi ^{\dag }\left( y\right) \delta \varphi \left(
x\right) }\right\vert _{f=0}.  \label{eq B.13}
\end{equation}
}


\section{ Finite Terms}

\label{sec:C}

Some of the resulting expression from the radiative corrections were too
lengthy to present them in the text body; so, we decided to present them
in the form of an appendix in order of complementarity.

\subsection{ Finite terms $\Sigma$}

\label{sec:C.1}

We have for the finite part of the meson self-energy the expression:{\small
\begin{eqnarray}
&&\Sigma _{fin}^{\left( 1\right) }\left( p\right) =\frac{\alpha }{4\pi }%
\int_{x}\left( \left( 4-26x+21x^{2}\right) p^{2}+2\left( 7x-4\right)
m^{2}\right)   \ln \left[ \frac{\Theta _{1}}{\Theta _{1}+\left( 1-x\right)
m_{P}^{2}}\right] \label{eq C.1}\\
&&+\frac{\alpha }{4\pi }\int_{x}\int_{w}\bigg\{\left[ \frac{\left( 1-\xi
\right) }{\Theta _{1}}-\frac{\xi }{\Theta _{1}+\left( 1-x\right) m_{P}^{2}}-%
\frac{\left( 1-2\xi \right) }{\Theta _{1}+wm_{P}^{2}}\right] \left( \left(
2-x^{2}\right) p^{2}+2m^{2}\right) x^{2}p^{2}  \notag \\
&&+\left( 2\left( 1-3x+6x^{2}\right) p^{2}+6wm_{P}^{2}-2\left( 2-3x\right)
m^{2}\right) \left[ \xi \ln \left[ \frac{\Theta _{1}+\left( 1-x\right)
m_{P}^{2}}{\Theta _{1}+wm_{P}^{2}}\right] +\left( 1-\xi \right) \ln \left[
\frac{\Theta _{1}+wm_{P}^{2}}{\Theta _{1}}\right] \right] \bigg\}. \notag
\end{eqnarray}
}With the following definitions:
\begin{equation}
\Theta _{1}=m^{2}x-x\left( 1-x\right) p^{2},
\end{equation}
{\small
\begin{equation}
\int_{x_{1}}\int_{x_{2}}\int_{x_{3}}...=\int_{0}^{1}dx_{1}%
\int_{0}^{1-x_{1}}dx_{2}\int_{0}^{1-x_{1}-x_{2}}dx_{3}...%
\end{equation}
}

\subsection{ Finite terms $\Lambda $}

 \label{sec:C.2}

Now, the finite parts of the vertex function $\Lambda$ are given by the following expressions:
{\small
\begin{eqnarray}
\Lambda _{\sigma }^{\left( a\right) }\left( p,s\right)&=&\frac{\alpha }{4\pi
}\int_{x}\int_{z}\bigg\{2\left[ \left( 2-3z\right) p+\left( 2-3x\right) s%
\right] _{\sigma }\ln \left[ \frac{\Theta _{2}+m_{P}^{2}\left( 1-x-z\right)
}{\Theta _{2}}\right] +\left[ \frac{1}{\Theta _{2}+m_{P}^{2}\left(
1-x-z\right) }-\frac{1}{\Theta _{2}}\right] A_{\sigma }\bigg\}  \notag \\
&&+\frac{\alpha }{4\pi }\int_{x}\int_{z}\int_{y}\left( \left( 1-2z\right)
p+\left( 1-2x\right) s\right) _{\sigma }\bigg\{\left[ \frac{\left( 1-2\xi
\right) }{\Theta _{2}+m_{P}^{2}y}-\frac{\left( 1-\xi \right) }{\left( \Theta
_{2}\right) }+\frac{\xi }{\left( \Theta _{2}+m_{P}^{2}\left( 1-x-z\right)
\right) }\right] P  \notag\\
&&+\left[ \frac{\left( 1-2\xi \right) }{\left( \Theta _{2}+m_{P}^{2}y\right)
^{2}}-\frac{\left( 1-\xi \right) }{\left( \Theta _{2}\right) ^{2}}+\frac{\xi
}{\left( \Theta _{2}+m_{P}^{2}\left( 1-x-z\right) \right) ^{2}}\right]
M\notag \\
&&+6\xi \ln \left[ \frac{\Theta _{2}+m_{P}^{2}y}{\Theta _{2}+m_{P}^{2}\left(
1-x-z\right) }\right] +6\left( 1-\xi \right) \ln \left[ \frac{\Theta _{2}}{%
\Theta _{2}+m_{P}^{2}y}\right] \bigg\},  \label{eq C.2} 
\end{eqnarray}
}and{\small
\begin{eqnarray}
\Lambda _{\sigma }^{\left( b\right) }\left( p\right)&=&\frac{\alpha }{2\pi }%
p_{\sigma }\int_{x}\left( 2-x\right) \ln \left[ \frac{\Theta _{1}}{\Theta
_{1}+\left( 1-x\right) m_{P}^{2}}\right]\notag \\
&& +\frac{\alpha }{4\pi }p_{\sigma }\int_{x}\int_{w}\bigg\{\left[ \frac{%
\left( 1-2\xi \right) }{\Theta _{1}+wm_{P}^{2}}-\frac{\left( 1-\xi \right) }{%
\Theta _{1}}+\frac{\xi }{\Theta _{1}+\left( 1-x\right) m_{P}^{2}}\right]
2\left( 2-x\right) x^{2}p^{2}  \notag \\
&&+2\left( 1-3x\right) \left[ \xi \ln \left[ \frac{\Theta _{1}+\left(
1-x\right) m_{P}^{2}}{\Theta _{1}+wm_{P}^{2}}\right] +\left( 1-\xi \right)
\ln \left[ \frac{\Theta _{1}+wm_{P}^{2}}{\Theta _{1}}\right] \right] \bigg\}; \label{eq C.3}
\end{eqnarray}
}with
\begin{eqnarray}
\Theta _{2}=m^{2}\left( x+z\right) -x\left( 1-x\right) s^{2}-z\left(1-z\right) p^{2}+xzs.p ,
\end{eqnarray}%
and
\begin{eqnarray}
A_{\sigma }=\left( xs-\left( 2-z\right) p\right) .\left( zp-\left(
2-x\right) s\right) \left( \left( 1-2z\right) p+\left( 1-2x\right) s\right)
_{\sigma } ,
\end{eqnarray}
and
\begin{eqnarray}
P &=&6z\left( 1-z\right) p^{2}+6x\left( 1-x\right) s^{2}-2\left(1-3x-3z+6xz\right) s.p \notag  \\
M &=&\left( z\left( 2-z\right) p^{2}-x^{2}s^{2}+2x\left( 1-z\right)
p.s\right) \left( x\left( 2-x\right) s^{2}-z^{2}p^{2}+2z\left( 1-x\right)
p.s\right).
\end{eqnarray}%

\subsection{Finite terms $\Psi $}

\label{sec:C.3}
At last, we have the following expressions for the finite parts of the vertex function $\Psi$:{\small
\begin{eqnarray}
\Psi _{\sigma \rho }^{\left( a\right) }& =&\frac{\alpha }{\pi }\eta _{\sigma
\rho }\int_{x}\ln \left[ \frac{\Theta _{3}+\left( 1-x\right) m_{P}^{2}}{%
\Theta _{3}}\right] \notag \\
&&+\frac{\alpha }{2\pi }\int_{x}\int_{w}\bigg\{\left[ \frac{\left( 1-\xi
\right) }{\Theta _{3}}-\frac{\xi }{\left( \Theta _{3}+\left( 1-x\right)
m_{P}^{2}\right) }-\frac{\left( 1-2\xi \right) }{\Theta _{3}+wm_{P}^{2}}%
\right] 2x^{2}\left( p+k\right) _{\sigma }\left( p+k\right) _{\rho } \notag \\
&& +\xi \eta _{\sigma \rho }\ln \left[ \frac{\Theta _{3}+wm_{P}^{2}}{\Theta
_{3}+\left( 1-x\right) m_{P}^{2}}\right] +\left( 1-\xi \right) \eta _{\sigma
\rho }\ln \left[ \frac{\Theta _{3}}{\Theta _{3}+wm_{P}^{2}}\right] \bigg\},\label{eq C.4}
\end{eqnarray}
}and{\small
\begin{eqnarray}
&&\Psi _{\sigma \rho }^{\left( b\right) } =\frac{\alpha }{2\pi }%
\int_{x}\int_{z}\bigg\{\left[ \frac{1}{\Theta _{4}}-\frac{1}{\left( \Theta
_{4}+\left( 1-x-z\right) m_{P}^{2}\right) }\right] A_{\sigma \rho }+\eta _{\sigma \rho }\ln \left[ \frac{\Theta _{4}}{\Theta _{4}+\left( 1-x-z\right)
m_{P}^{2}}\right] \bigg\}  \notag \\
&&+\frac{\alpha }{4\pi }\int_{x}\int_{z}\int_{y}\bigg\{6\eta _{\sigma \rho }\left( 1-\xi \right) \ln \left[ \frac{\Theta
_{4}+ym_{P}^{2}}{\Theta _{4}}\right]+\bigg[ \frac{\xi }{
\left( \Theta _{4}+\left( 1-x-z\right) m_{P}^{2}\right) }+\frac{\left(
1-2\xi \right) }{\left( \Theta _{4}+ym_{P}^{2}\right) } -\frac{\left( 1-\xi
\right) }{\Theta _{4}}\bigg] B_{\sigma \rho }  \notag \\
&&+\left[ \frac{\left( 1-\xi \right) }{\left( \Theta _{4}\right) ^{2}}-\frac{%
\left( 1-2\xi \right) }{\left( \Theta _{4}+ym_{P}^{2}\right) ^{2}}-\frac{\xi
}{\left( \Theta _{4}+\left( 1-x-z\right) m_{P}^{2}\right) ^{2}}\right]
2C_{\sigma \rho }  +6\eta _{\sigma \rho }\xi \ln \left[
\frac{\Theta _{4}+\left( 1-x-z\right) m_{P}^{2}}{\Theta _{4}+ym_{P}^{2}}%
\right] \bigg\},  \label{eq C.5}
\end{eqnarray}
}also{\small
\begin{eqnarray}
&&\Psi _{\sigma \rho }^{\left( d\right) } =\frac{\alpha }{8\pi }%
\int_{x}\int_{z}\int_{w}\bigg\{\left[ \frac{1}{\left( \Theta _{5}\right) ^{2}%
}-\frac{1}{\left( \Theta _{5}+\left( 1-x-z-w\right) m_{P}^{2}\right) ^{2}}%
\right] 2K_{\sigma \rho }  \label{eq C.6} \\
&&+\left[ \frac{1}{\left( \Theta _{5}+\left(
1-x-z-w\right) m_{P}^{2}\right) }-\frac{1}{\Theta _{5}}\right] D_{\sigma
\rho }+12\eta _{\sigma \rho }\ln \left[ \frac{\Theta _{5}+\left( 1-x-z-w\right)
m_{P}^{2}}{\Theta _{5}}\right] \bigg\}\notag \\
&&+\frac{\alpha }{16\pi }%
\int_{x}\int_{z}\int_{w}\int_{t}\bigg\{\left[ \frac{\left( 1-\xi \right) }{\left( \Theta _{5}\right) }-\frac{%
\left( 1-2\xi \right) }{\left( \Theta _{5}+ym_{P}^{2}\right) }-\frac{\xi }{%
\left( \Theta _{5}+\left( 1-x-z-w\right) m_{P}^{2}\right) }\right] G_{\sigma
\rho }    \notag \\
&&+64\eta _{\sigma \rho }\left( 1-\xi \right) \ln \left[ \frac{\Theta _{5}}{%
\Theta _{5}+ym_{P}^{2}}\right] +\left[ \frac{\left( 1-\xi \right) }{\left(
\Theta _{5}\right) ^{3}}-\frac{\xi }{\left( \Theta _{5}+\left(
1-x-z-w\right) m_{P}^{2}\right) ^{3}}-\frac{\left( 1-2\xi \right) }{\left(
\Theta _{5}+ym_{P}^{2}\right) ^{3}}\right] 8E_{\sigma \rho }  \notag \\
&&+\left[ \frac{\xi }{\left( \Theta _{5}+\left( 1-x-z-w\right)
m_{P}^{2}\right) ^{2}}-\frac{\left( 1-\xi \right) }{\left( \Theta
_{5}\right) ^{2}}+\frac{\left( 1-2\xi \right) }{\left( \Theta
_{5}+ym_{P}^{2}\right) ^{2}}\right] 2F_{\sigma \rho }  +64\eta _{\sigma \rho }\xi \ln \left[
\frac{\Theta _{5}+ym_{P}^{2}}{\Theta _{5}+\left( 1-x-z-w\right) m_{P}^{2}}%
\right]\bigg\},  \notag
\end{eqnarray}
}and, at last{\small
\begin{eqnarray}
\Psi _{\sigma \rho }^{\left( e\right) }&=&\frac{\alpha }{2\pi }\eta _{\sigma
\rho }\int_{x}\int_{z}\bigg\{\left( \left( 2-z\right) p-xs\right) .\left(
\left( 2-x\right) s-zp\right) \left[ \frac{1}{\Theta _{2}}-\frac{1}{\Theta
_{2}+\left( 1-x-z\right) m_{P}^{2}}\right]+2\ln \left[ \frac{\Theta _{2}+\left( 1-x-z\right) m_{P}^{2}}{\Theta _{2}}%
\right] \bigg\}\notag \\
&&+\frac{\alpha }{2\pi }\eta _{\sigma \rho
}\int_{x}\int_{z}\int_{y}\bigg\{\left[ \frac{\left( 1-\xi \right) }{\left(
\Theta _{2}\right) ^{2}}-\frac{\xi }{\left( \Theta _{2}+\left( 1-x-z\right)
m_{P}^{2}\right) ^{2}}-\frac{\left( 1-2\xi \right) }{\left( \Theta
_{2}+ym_{P}^{2}\right) ^{2}}\right] K  \notag \\
&&+\left[ \frac{\left( 1-\xi \right) }{\Theta _{2}}-\frac{\xi }{\Theta
_{2}+\left( 1-x-z\right) m_{P}^{2}}-\frac{\left( 1-2\xi \right) }{\left(
\Theta _{2}+ym_{P}^{2}\right) }\right] P\notag \\
&&+6\xi \ln \left[ \frac{\Theta
_{2}+\left( 1-x-z\right) m_{P}^{2}}{\Theta _{2}+ym_{P}^{2}}\right] +6\left(
1-\xi \right) \ln \left[ \frac{\Theta _{2}+ym_{P}^{2}}{\Theta _{2}}\right] %
\bigg\}; \label{eq C.7}
\end{eqnarray}
}where
\begin{eqnarray}
\Theta _{3}&=&m^{2}x-x\left( 1-x\right) \left( p+k\right) ^{2}, \notag\\
\Theta _{4}&=&m^{2}\left( x+z\right) +2xzs\left( p+k\right) -x\left(
1-x\right) s^{2}-z\left( 1-z\right) \left( p+k\right) ^{2}, \notag\\
\Theta _{5}&=&m^{2}\left( x+z+w\right) -x\left( p+k\right)
^{2}-zs^{2}-wp^{2}+\left( xk+zs+\left( x+w\right) p\right) ^{2},
\end{eqnarray}%
and
\begin{eqnarray}
K =\left( \left( 2-z\right) p-xs\right) .\left( zp+xs\right) \left( \left(
2-x\right) s-zp\right) .\left( zp+xs\right) .
\end{eqnarray}%
Also the polynomials:{\small
\begin{eqnarray}
A_{\sigma \rho }&=&\left( 1-2z\right) \left( 2-x\right) \left( p+k\right)
_{\rho }s_{\sigma }-z\left( 1-2x\right) s_{\rho }\left( p+k\right) _{\sigma }+\left( 1-2x\right) \left( 2-x\right) s_{\sigma }s_{\rho
}\notag \\
&&-z\left( 1-2z\right) \left( p+k\right) _{\sigma }\left(
p+k\right) _{\rho }; \notag \\
B_{\sigma \rho } &=&2\left( 1-2z-3x+8xz\right) s_{\sigma }\left( p+k\right)
_{\rho }+2z\left( 8x-5\right) \left( p+k\right) _{\sigma }s_{\rho }+2\left( 1-7x+8x^{2}\right) s_{\sigma }s_{\rho }\notag \\
&&+2z\left( 8z-3\right) \left( p+k\right) _{\sigma }\left( p+k\right)
_{\rho }-2\eta _{\sigma
\rho }\left( \left( 2-x\right) s-z\left( p+k\right) \right) .\left(
xs+z\left( p+k\right) \right) ; \notag \\
C_{\sigma \rho }&=&\left( \left( 2-x\right) s-z\left( p+k\right) \right) .\left( xs+z\left(
p+k\right) \right)\left( \left( 1-2z\right) \left( p+k\right) +\left(
1-2x\right) s\right) _{\rho }\left( xs+z\left( p+k\right) \right) _{\sigma
} ;
\end{eqnarray}%
}and{\small
\begin{eqnarray}
D_{\sigma \rho } &=&4\left( 2\left( 1-x-w\right) p+\left( 1-2x\right)
k-2zs\right) _{\sigma }\left( \left( 2-3x-3w\right) p+\left( 1-3x\right)
k+\left( 2-3z\right) s\right) _{\rho } \notag \\
&&+4\left( \left( 1-2x-2w\right) p+\left( 1-2x\right) k+\left( 1-2z\right)
s\right) _{\rho } \left( \left( 1-z\right) s-xk+\left( 1-x-w\right) p\right)
_{\sigma }  \notag \\
&&+4\eta _{\sigma \rho }\left( \left( 2-z\right) s-xk-\left( x+w\right)
p\right) .\left( \left( 2-x-w\right) p-xk-zs\right) ; \notag \\
E_{\sigma \rho } &=&\left( 2\left( 1-x-w\right) p+\left( 1-2x\right)
k-2zs\right) _{\sigma }\left( \left( 1-2x-2w\right) p+\left( 1-2x\right)
k+\left( 1-2z\right) s\right) _{\rho } \notag \\
&&\times \left( \left( 2-x-w\right) p-xk-zs\right) .\left( xk+zs+\left(
x+w\right) p\right)\notag \\
&& \times \left( \left( 2-z\right) s-xk-\left( x+w\right) p\right)
.\left( xk+zs+\left( x+w\right) p\right) ; \notag \\
G_{\sigma \rho } &=&24\left( 2\left( 1-x-w\right) p+\left( 1-2x\right)
k-2zs\right) _{\sigma } \left( 2\left( 1-2x-2w\right) p+\left( 1-4x\right)
k+2\left( 1-2z\right) s\right) _{\rho }  \notag \\
&&+24\left( \left( 1-2x-2w\right) p+\left( 1-2x\right) k+\left( 1-2z\right)
s\right) _{\rho }\left( \left( 1-2x-2w\right) p-2xk+\left( 1-2z\right)
s\right) _{\sigma }  \notag \\
&&-48\eta _{\sigma \rho }\left( \left( 1-x-w\right) p-xk+\left( 1-z\right)
s\right) .\left( xk+zs+\left( x+w\right) p\right) \notag \\
&&+16\left( \left( 1-x-w\right) p-xk-zs\right) ^{\alpha }\left( \left(
1-z\right) s-xk-\left( x+w\right) p\right) ^{\beta } \left( \eta _{\alpha
\beta }\eta _{\sigma \rho }+\eta _{\alpha \sigma }\eta _{\beta \rho }+\eta
_{\alpha \rho }\eta _{\beta \sigma }\right); \notag \\
K_{\sigma \rho } &=&\left( 2\left( 1-x-w\right) p+\left( 1-2x\right)
k-2zs\right) _{\sigma } \left( \left( 1-2x-2w\right) p+\left( 1-2x\right)
k+\left( 1-2z\right) s\right) _{\rho } \notag \\
&&\times \left( \left( 2-z\right) s-xk-\left( x+w\right) p\right)  .\left(\left( 2-x-w\right) p-xk-zs\right) .
\end{eqnarray}%
}And{\small
\begin{eqnarray}
F_{\sigma \rho } &=&-8\left( 2\left( 1-x-w\right) p+\left( 1-2x\right)
k-2zs\right) _{\sigma }\left( \left( 1-2x-2w\right) p+\left( 1-2x\right)
k+\left( 1-2z\right) s\right) _{\rho } \notag \\
&& \times \left( \left( 1-x-w\right) p-xk+\left(
1-z\right) s\right) .\left( xk+zs+\left( x+w\right) p\right)   \notag \\
&&+4\left( 2\left( 1-x-w\right) p+\left( 1-2x\right) k-2zs\right) _{\sigma
}\left( \left( 1-2x-2w\right) p+\left( 1-2x\right) k+\left( 1-2z\right)
s\right) _{\rho } \notag \\
&& \times \left( \left( 1-x-w\right) p-xk-zs\right) .\left( \left(
1-z\right) s-xk-\left( x+w\right) p\right)   \notag \\
&&+4\left( 2\left( 1-x-w\right) p+\left( 1-2x\right) k-2zs\right) _{\sigma
}\left( xk+\left( x+w\right) p+zs\right) _{\rho } \notag \\
&&\times\left( \left( 2-x-w\right)
p-xk-zs\right) .\left( xk+zs+\left( x+w\right) p\right)   \notag \\
&&+4\left( 2\left( 1-x-w\right) p+\left( 1-2x\right) k-2zs\right) _{\sigma
}\left( xk+\left( x+w-1\right) p+\left( z-1\right) s\right) _{\rho }   \notag \\
&&\times \left(\left( 2-z\right) s-xk-\left( x+w\right) p\right) .\left( xk+zs+\left(
x+w\right) p\right)   \notag \\
&&-4\left( \left( 1-2x-2w\right) p+\left( 1-2x\right) k+\left( 1-2z\right)
s\right) _{\rho }\left( \left( 1-z\right) s-xk-\left( x+w\right) p\right)
_{\sigma } \notag \\
&&\times \left( \left( 2-x-w\right) p-xk-zs\right) .\left( xk+zs+\left(
x+w\right) p\right)   \notag \\
&&+4\left( \left( 1-2x-2w\right) p+\left( 1-2x\right) k+\left( 1-2z\right)
s\right) _{\rho }\left( xk+zs+\left( x+w-1\right) p\right) _{\sigma } \notag \\
&&\times \left( \left( 2-z\right) s-xk-\left( x+w\right) p\right) .\left( xk+zs+\left(
x+w\right) p\right)   \notag \\
&&+4\eta _{\sigma \rho }\left( \left( 2-x-w\right) p-xk-zs\right) .\left(
xk+zs+\left( x+w\right) p\right) \notag \\
&&\times\left( \left( 2-z\right) s-xk-\left( x+w\right) p\right) .\left( xk+zs+\left( x+w\right) p\right) .
\end{eqnarray}
}


\end{document}